\documentclass[our,fleqn]{w-art}
\usepackage{times}
\usepackage{amsmath}
\usepackage{amssymb}
\usepackage{latexsym}
\usepackage{wasysym}

\begin{document}
\title[Radion-induced
  gravitational wave oscillations]{Radion-induced gravitational wave oscillations and their phenomenology}
\author[A.\ O.\ Barvinsky]{Andrei O.\ Barvinsky\footnote{
e-mail: barvin@td.lpi.ru}\inst{1}} 
\author[A.\ Yu.\  Kamenshchik]{Alexander Yu.\  Kamenshchik\footnote{
e-mail: sasha.kamenshchik@centrovolta.it}\inst{2,3}}  
\author[A.\  Rathke]{Andreas  Rathke\footnote{
e-mail: andreas.rathke@physik.uni-freiburg.de }\inst{4,5}}
\author[C.\ Kiefer]{Claus Kiefer\footnote{
e-mail: kiefer@thp.uni-koeln.de}\inst{5}}
\address[\inst{1}]{Theory Department, Lebedev Physics Institute,
Leninsky Pr.\ 53,
Moscow 117924, Russia}
\address[\inst{2}]{L.\ D.\ Landau Institute for Theoretical Physics of the Russian
Academy of Sciences, Kosygina street\ 2, Moscow 117334, Russia}
\address[\inst{3}]{Landau Network --- Centro Volta, Villa Olmo, via Cantoni 1,
  22100 Como, Italy}
\address[\inst{4}]{Physikalisches Institut, Universit\"at Freiburg,
  Hermann-Herder-Stra\ss e 3, 79104 Freiburg, Germany}
\address[\inst{5}]{Institut f\"ur Theoretische Physik, 
Universit\"at zu K\"oln, 
Z\"ulpicher Stra\ss e 77, 50937 K\"oln, Germany}
\maketitle
\begin{abstract}
  We discuss the theory and phenomenology of the interplay between the
  massless graviton and its massive Kaluza-Klein modes in the Randall-Sundrum
  two-brane model.  The equations of motion of the transverse traceless
  degrees of freedom are derived by means of a 
  Green function approach as well as from
  an effective nonlocal action.  The second procedure clarifies the
  extraction of the particle content from the nonlocal action and the issue of
  its diagonalization. The situation discussed is generic for the treatment of
  two-brane models if the on-brane fields are used as the dynamical degrees of
  freedom. The mixing of the effective graviton modes of the localized action
  can be interpreted as radion-induced gravitational-wave oscillations, a
  classical analogy to meson and neutrino oscillations.  We show that these
  oscillations arising in M-theory-motivated braneworld setups could lead to
  effects detectable by gravitational-wave interferometers. The implications
  of this effect for models with ultra-light gravitons are discussed.

\end{abstract}

\section{Introduction\label{intro}}

In recent years the old Kaluza-Klein idea of a world with additional
dimensions got a strong impact from attempts motivated by string theory
to resolve the hierarchy problem \cite{kubyshinrubakov}. Especially fruitful
is the idea of implementing warped manifolds, for which the metric of the
four-dimensional space-time can depend on the additional dimensions.  The most
popular models using warped geometries are the Randall-Sundrum (RS) models
\cite{RS1,RS2} in which our observable Universe is treated as a
four-dimensional brane embedded into a five-dimensional anti-de~Sitter (AdS)
bulk space.  The particular properties of the AdS geometry allow to describe
the trapping of the gravitational zero-mode on the four-dimensional brane
\cite{RS2}.  Displacing the second four-dimensional brane in the bulk, one can
propose a resolution to the hierarchy problem \cite{RS1}.

However, warped two-brane models have some important additional features.  The
most interesting effect is the existence of {\em light} massive graviton modes,
which have already attracted a lot of attention in the context of higher-spin
theories and in a cosmological context \cite{Vas,Deser}. A particularly
intriguing development is the construction of semi-realistic braneworld models
in which the Kaluza-Klein gravitons\footnote{In the following we will denote
  also a {\em classical} linearized gravitational mode by the term graviton.}
(KK gravitons) can have an extremely low mass and can be responsible for the
mediation of all \cite{GRS,KR} or a large part of the gravitational interaction
\cite{bigrav,KMP}.  In these setups, the Veltman-van-Dam-Zakharov discontinuity
(VvDZ discontinuity) \cite{VvDZ} (for a recent analysis of VvDZ discontinuity
at the classical level see \cite{CarGiu})  is either unobservable due to the
setup \cite{KRVvDZ} or cured by non-linear effects \cite{cure}, though it
is still under dispute if these models exhibit a realistic behavior in the
limit of strong gravitational fields \cite{dispute}.

An effect so far overlooked is the mixing between graviton modes of different
masses which inevitably occurs in higher-dimensional spacetime models. This
effect is of little relevance in models with small extra dimensions and heavy
KK modes, because in these models the massive gravitons will rarely be produced
and therefore are of no astrophysical significance. In models where gravitons
can have masses in the sub-eV scale like in the RS models, the mixing between
the different modes can lead to interesting effects in gravitational-wave
astrophysics. Gravitational waves (GW's) will then consist of contributions
from several graviton modes. This will induce a beat in the gravitational wave
which could be detectable with gravitational-wave interferometers. The
parameters of the oscillations will crucially depend on the size of the extra
dimension --- commonly parameterized by the radion-field ---, thus suggesting
the notion of radion-induced graviton oscillations (RIGO's).  This effect was
investigated in \cite{rigoshort} and will be discussed in full detail in the
present paper.

RIGO's are particularly interesting because their observability is not limited
to a certain minimal value of the extra-dimensional curvature scale.  RIGO's
could therefore become the first detectable signature of M-theory
phenomenology.  From the very beginning, the affinity of the RS two-brane setup
to the M-theory compactification of \cite{LOSW} (see \cite{Mcosmo} for reviews
of effective five-dimensional M-theory) was noticed \cite{RS1}. Actually, the
RS two-brane model can be viewed as a simplified version of this M-theory
solution which appears five-dimensional in an intermediate energy-range. The
RS model simplifies the effective supergravity action of \cite{LOSW} to an
Einstein-Hilbert action and neglects the gauge-field part of the M-theory
effective action. Due to this relationship, the RS model has been used as an
approximation for aspects of M-theory cosmology where a full-fledged M-theory
calculation is not feasible (see \cite{Ekpyr} for an example). 

On the other hand, the RS model deviates in major aspects from the requirements
of M-theory: First, in contrast to the RS model, M-theory restricts the size
of the largest extra dimension to typically $10^{-31}\,{\rm m} \le l \le
10^{-29}\,{\rm m}$ \cite{WBD,Mcosmo} (see, however, \cite{Lykken}). This rules
out the possibility to detect the extra dimension by measuring short-distance
deviations from Newtonian gravity. Second, in M-theory the requirements of
having the standard-model group structure and supersymmetry breaking forces
one to identify our Universe with the positive-tension brane (cf.\ 
\cite{Pyr}).  This prevents one from obtaining the simple geometrical solution
to the hierarchy problem by locating our Universe on the 
negative tension brane of
the RS model.\footnote{There may remain some loopholes though, see for
  example \cite{PomarolGNS} for a construction which would allow us to live
  on the positive-tension brane and have the hierarchy inherited from the
  negative-tension brane or \cite{Benakli,Donagi} for M-theory constructions
  which would allow the standard-model fields to reside on the
  negative-tension brane.}
Therefore it seems that one has to choose between either the attractive
phenomenology of the RS model or the appealing theoretical framework of
M-theory, although it remains tempting to conjecture that if M-theory realizes
large extra-dimensions in nature they should exhibit a similar setup as the
Randall-Sundrum two-brane model.

Considering RIGO's one may, however, be able to probe several additional
orders of smallness of the extra-dimensional curvature radius if only the
radion acquires suitable values. Although the required values of the radion
are still far from the parameter range of typical M-theory models, the search
for RIGO's thus offers a possibility to narrow down the gap between the
phenomenologically accessible parameter space of large extra dimensions and
the predictions of M-theory.

In our preceding paper \cite{we} we have studied the four-dimensional
effective action in the RS two-brane model.  This action is a functional of
the two induced metrics and radion fields on the two four-dimensional branes.
Using the results of \cite{rigoshort,rigosakh}, we shall here extend the study
of radion-induced graviton oscillations in the two-brane world of
\cite{rigoshort} and investigate their phenomenological consequences.

The structure of this paper is the following. In the next section we will
introduce the RS setup and describe the corresponding nonlocal braneworld
effective action and the equations of motion in terms of a nonlocal kernel and
Green function, respectively. In Sec.\ \ref{eigen} to \ref{diag} we will then
set up the theoretical framework for studying RIGO's by deriving the localized
equations of motion from the nonlocal theory. In Sec.~\ref{eigen} we will
present an explicit expansion of the Green function in terms of its residues
which leads to the equations of motion in terms of orthogonal mass eigenmodes
of gravitons.  In Sec.~\ref{actionsec} we will demonstrate how these modes can
be extracted directly from the action. We find that there are two sets of
eigenmodes of the action and explain how they are interrelated.
To further elucidate this finding,
we undertake in Sec.~\ref{diag} a direct diagonalization of the
action in a limit in which the two sets of modes coincide.  This method was
first employed in the preliminary study of RIGO's in \cite{rigosakh}.  The
rest of the paper will be devoted to phenomenology: In Sec.~\ref{phen} we
discuss the description of RIGO's and their possible observational
consequences in the RS model.  After reminding the reader of the analogous
effect of quantum oscillations in Sec.\ \ref{quantum}, we first discuss the
phenomenology of RIGO's on the positive-tension brane in Sec.\ \ref{RIGO+}.
This will lead us, in Sec.\ \ref{Mtheory}, 
to consider in particular the gravitational waves produced
by a network of cosmic strings on the negative-tension brane.
In Sec.\ \ref{RIGO-} follows the discussion of RIGO's on the
negative-tension brane. Finally, in Sec.\ \ref{bigrav}, the discussion is
extended to so-called bi-gravity braneworld models. In the concluding Sec.\ 
\ref{conc}, we summarize our results and argue for the occurrence of RIGO's in
a general class of braneworlds featuring orbifolded compactifications. A
detailed presentation of the diagonalization procedure for the kinetic and
massive parts of the gravitational effective action used in Sec.~\ref{diag} is
given in the Appendix.

\section{The RS two-brane model and the braneworld effective action\label{setup}}
 
In the RS setup \cite{RS1}, the braneworld effective action is induced by
the five-dimensional bulk spacetime as follows. We start with the theory 
having the action of the five-dimensional gravitational field with metric 
$G=G_{AB}(x,y)$, $A=(\mu,5),\,\mu=0,1,2,3$, propagating in the bulk spacetime 
$x^A=(x,y),\,x=x^\mu,\,x^5=y$, and matter fields $\phi$ confined to the two 
branes $\Sigma_\pm$, 
\begin{equation}
S[\,G,g,\phi\,]=S_5[\,G\,] + S_4[\,G,g,\phi\,], \label{action}
\end{equation}
where the bulk-space part of the action is given by
\begin{equation}
     S_5[\,G\,]=\frac1{16\pi G_5} 
     \int\limits_{M^5} d^5x\,G^{1/2} 
     \left(\,\vphantom{I}^5\!R(G)-2\Lambda_5\right),  \label{action1} 
\end{equation} 
and the brane-part of the action is
\begin{equation} 
     S_4[\,G,g,\phi\,]=
     \sum\limits_{\pm}\int_{\Sigma_\pm}\! 
     d^4x\,\Big( L_m(\phi,\partial\phi,g)
     -g^{1/2}\sigma_\pm
     + \frac1{8\pi G_5}[K]\Big). \label{actionbr}
\end{equation}

     The branes are marked by the index $\pm$ and carry induced metrics
     $g=g_{\mu\nu}(x)$ and matter-field Lagrangians
     $L_m(\phi,\partial\phi,g)$. The bulk part of the action (\ref{action1})
     is characterized by the five-dimensional gravitational constant $G_5$ and
     the cosmological constant $\Lambda_5$, while the brane parts
     (\ref{actionbr}) carry four-dimensional cosmological constants
     $\sigma_\pm$.  The bulk cosmological constant $\Lambda_5$ is negative
     and, thus, is capable of generating an AdS geometry, while the brane
     cosmological constants play the role of brane tensions and, depending on
     the model, can be of either sign. The Einstein-Hilbert bulk action
     (\ref{action1}) is accompanied by terms in (\ref{actionbr}) containing
     the jumps of extrinsic curvatures traces $[K]$ associated with both sides
     of each brane \cite{ChR}.  Solving the five-dimensional Einstein
     equations with prescribed values of the four-dimensional metric on the
     branes, we have obtained, in \cite{we}, the tree-level effective
     four-dimensional action.
 
In the RS two-brane setup, the fifth dimension has the topology of a circle
labeled by the coordinate $y$, $-d\leq y\leq d$, with an orbifold ${\mathbb
  Z}_2$-identification of points $y$ and $-y$. The branes are located at
antipodal fixed points of the orbifold, $y=y_\pm,\,y_+=0,\,|y_-|=d$. They are
empty, i.\,e.\ $L_m(\phi,\partial\phi,g_{\mu\nu})=0$, and their tensions are
opposite in sign and fine-tuned to the values of $\Lambda_5$ and $G_5$,
     \begin{equation} 
     \Lambda_5=- \frac 6 {l^2},\quad\quad 
     \sigma_+=-\sigma_-=\frac 3 {4\pi G_5l}.   \label{RS1} 
     \end{equation} 
Then this model admits a solution to the Einstein equations with an AdS 
metric in the bulk ($l$ is its curvature radius), 
     \begin{eqnarray} 
     ds^2=dy^2+e^{-2|y|/l}\eta_{\mu\nu}dx^\mu dx^\nu,  \label{RS2} 
     \end{eqnarray} 
$0=y_+\leq|y|\leq y_-=d$, and with a flat induced metric 
$\eta_{\mu\nu}$ on both branes \cite{RS1}. With the fine tuning 
(\ref{RS1}), this solution exists for arbitrary brane separation 
$d$. 

Now consider small matter sources
and metric perturbations $\gamma_{AB}(x,y)$ on the background of this
solution \cite{RS2,GT,GKRChGR},
     \begin{eqnarray}
     ds^2=dy^2+e^{-2|y|/l}\eta_{\mu\nu}dx^\mu dx^\nu
     +\gamma_{AB}(x,y)\,dx^Adx^B.                               \label{metric}
     \end{eqnarray}
Then this five-dimensional metric {\em induces} on the branes two
four-dimensional metrics of the form
    \begin{eqnarray}
    g^\pm_{\mu\nu}(x)=
    a^2_\pm\,\eta_{\mu\nu}+\gamma^\pm_{\mu\nu}(x),  \label{metric1}
    \end{eqnarray}
where the scale factors $a_\pm=a(y_\pm)$ are given by
\begin{equation}
    a_+=1, \quad a_-=e^{-d/l}\equiv a, \label{a-factor}    
\end{equation}
and $\gamma^\pm_{\mu\nu}(x)$ are the perturbations by which the brane metrics
$g^\pm_{\mu\nu}(x)$ differ from the (conformally) flat metrics $a_\pm^2
\eta_{\mu\nu}$ of the RS solution (\ref{RS2}). The variable $a(y)$ represents
the `modulus' --- the global homogeneous part of the radion field determining
the interbrane separation.

Due to the metric perturbations, the branes no longer stay at fixed values of
the fifth coordinate $y$.  Up to four-dimensional diffeomorphisms, their
embedding variables consist of two four-dimensional scalar fields, the
{\em linearized} radions $\psi^\pm(x)$, and the braneworld action can depend on
these scalars. Their geometrically invariant meaning is revealed in a special
coordinate system where the bulk-metric perturbations $\gamma_{AB}(x,y)$
satisfy the ``Randall-Sundrum gauge conditions'' $\gamma_{A5}=0$,
${\gamma_{\mu\nu}}^{,\,\nu}=h_{\ \mu}^\mu=0$. In this coordinate system, the brane
embeddings are defined by the equations
   \begin{eqnarray}
   \Sigma_\pm:\,\,\,
   y=y_\pm+\frac l{a^2_\pm}\psi^\pm(x),\quad \quad
   y_+=0,\quad y_-=d.                          \label{embed}
   \end{eqnarray}
    Therefore the radions describe the bending of the branes in the bulk (cf.\ 
   \cite{GT,GKRChGR}). 
   
   In \cite{we} we have derived the effective braneworld action in terms of
   the four-dimensional on-brane metrics (\ref{metric1}) and nonlocal
   matrix-valued form factors. We call the part of the action quadratic in
   $h^\pm_{\mu\nu}(x)$ --- the transverse-traceless parts of the full metric
   perturbations $\gamma_{\mu\nu}^\pm(x)$ on the branes --- the {\em graviton
     sector}.  It can be written as the following $2\times2$ quadratic form
   in terms of the metric perturbations and the special nonlocal operator
   ${\bf F}(\Box)$,
\begin{equation}  
    S_{\rm grav}[\,h^\pm_{\mu\nu}\,]  
    =\frac1{16\pi G_4}\int  
    d^4x\,\frac12\,  
    [h^+_{\mu\nu} \, h^-_{\mu\nu}]  
    \,\frac{{\bf  F}(\Box)}{l^2}\,  
\left[\begin{array}{c}  
    \,\,h_+^{\mu\nu} \\  
    \,\,h_-^{\mu\nu} \,  
    \end{array}\right]\,.        \label{4.24}  
\end{equation} 
 The effective gravitational constant $G_4$ is given by $G_4
\equiv G_5/l$. As was shown in Ref.\ \cite{we}, the operator ${\bf F}(\Box)$
is a complicated non-linear function of the D'Alembert operator $\Box$
expressed by means of Bessel and Neumann functions of the arguments $l
\sqrt{\Box}$ and $l\sqrt{\Box}/a$,
\begin{equation}
    {\bf F}(\Box) =-\frac{1}{ J_2^+\,Y_2^- -J_2^-\,Y_2^+}
    \left[\begin{array}{cc}
    \,l \sqrt{\Box}\, u_+(l\sqrt{\Box}/a)&-\displaystyle{\frac2\pi} \\
    -\displaystyle{\frac2\pi}& \frac{l \sqrt{\Box}}a\, u_-(l\sqrt{\Box})
    \end{array}\right]\, ,                                 \label{4.11}
\end{equation}
where
\begin{align}
    &u_\pm(z)=Y_1^\pm J_2(z)
    -J_1^\pm Y_2(z),& &                     \label{4.6}\\
    &J_1^+ \equiv J_1(l\sqrt\Box),\,\,\,\,
    &Y_1^+ &\equiv Y_1(l\sqrt\Box),& &\\
    &J_1^- \equiv J_1(l\sqrt\Box /a),\,\,\,\,
    &Y_1^- &\equiv Y_1(l\sqrt\Box /a).& & \label{4.6b}
\end{align}
The action (\ref{4.24}) should be amended by the standard coupling of the
transverse-traceless gravitational modes to the {\em transverse-traceless}
part of the stress-energy tensors, $T^\pm_{\mu\nu}$, on the two branes,
\begin{equation}
S_\text{mat} =
\int d^4x [h^+_{\mu\nu},h^-_{\mu\nu}] 
\left[\begin{array}{c}T^+_{\mu\nu}\\T^-_{\mu\nu}\end{array}\right] \, .
\label{matteraction}
\end{equation}
The inverse of the kernel ${\bf F}(\Box)$ is the Green
function ${\bf G(\Box)}$ of the problem,
\begin{equation}
{\bf G} \equiv {\bf F}^{-1}.
\end{equation}
With the abbreviations given in (\ref{4.6}) -- (\ref{4.6b}), it reads
\begin{equation}
    {\bf G}(\Box) =\frac{a}{l^2 \Box}
    \frac{1}{J_1^+\,Y_1^-  - J_1^-\,Y_1^+ }
    \left[\begin{array}{cc}
    \,\displaystyle{\frac{l \sqrt{\Box}}a} \,u_-(l\sqrt\Box)
    &\displaystyle{\frac2\pi} \\
    \displaystyle{\frac2\pi}&l \sqrt{\Box}\, u_+(l\sqrt\Box/a)\,
    \end{array}\right]\, .                               \label{4.10}
\end{equation}
It has first been calculated in
\cite{Grinstein}.
With the help of ${\bf G(\Box)}$, one
finds the equations of motion for the transverse-traceless sector
\begin{eqnarray}
\left[\begin{array}{c} h^+_{\mu\nu}\\ h^-_{\mu\nu} \end{array} \right]
= - 8 \pi G_4 l^2\,{\bf G}(\Box)
    \left[\begin{array}{c} 
T^+_{\mu \nu} \\ T^-_{\mu\nu} \end{array}
\right],
  \label{eom}
\end{eqnarray}
corresponding to the variation of the combined action of (\ref{4.24}) and
(\ref{matteraction}). In this paper, we study in detail the properties of the
model in the \emph{low-energy limit} when $l \sqrt{\Box} \ll 1$ but when $l
\sqrt{\Box}/a$ can take arbitrary values in view of the smallness of the
parameter $a=e^{-d/l}$ (large interbrane distance).

As we will only be considering the transverse-traceless part of the
gravitational dynamics, tensor indices will be omitted for the rest of this  
paper.

%
%

\section{The eigenmode expansion of the Green function \label{eigen}}

Both the Green function ${\bf G}(\Box)$ and the kernel of the action ${\bf
  F}(\Box)$ are highly nonlocal. Nevertheless, it is possible to obtain a
conventional interpretation of the equations of motion and the action in
terms of an infinite tower of orthogonal Kaluza-Klein modes. We will first
discuss the recovery of the KK tower from the Green function because this
procedure works in a more direct way than the recovery of the particle
spectrum from the action, which will be discussed in the Sec.\ \ref{actionsec}.
The elements $G_{\times\times}$ ($\times = \pm$) of ${\bf G}(\Box)$ are
meromorphic functions of $\Box$. If we consider a concentric circle $C_n$
around $\Box = 0$ so that $C_n$ includes the first $n$ poles and let its
radius $R_n \to \infty$ as $n \to \infty$, we find the falloff property
\begin{equation}
|G_{\times\times}(\Box)| < \epsilon R_n\, ,
\quad\quad \quad\quad
R_n \to \infty\, , 
\label{falloff}
\end{equation}
for any small constant $\epsilon$ and all $\Box$ on $C_n$. Therefore we can
apply the Mittag-Leffler expansion, which provides an expansion of a
meromorphic function in terms of its poles, to the Green function 
in order to obtain a
representation of ${\bf G}(\Box)$ as a sum of scalar propagators.

In the limit $l\sqrt\Box \ll 1$ but $l\sqrt\Box/a$ arbitrary, the Green
function ${\bf G}(\Box)$ is given by\footnote{Note that the falloff property
  (\ref{falloff}) is not valid for the element $G_{++}$ in the approximation
  (\ref{Gsmalla}) but only for the full expression (\ref{4.10}). This does of
  course not prevent us to use the approximation (\ref{Gsmalla}) for an
  approximate determination of the residues and thus for an approximate
  determination of the Mittag-Leffler expansion.}
\begin{equation}
{\bf G}(\Box) \approx
\left[
\begin{array}{cc}
\frac{2}{l^2\Box}-\frac 12 
-\frac\pi 2 \displaystyle{\frac{Y_1[l\sqrt{\Box}/a]}{J_1[l\sqrt{\Box}/a]}}
+ {\bf C} + \ln\left( \frac{l\sqrt{\Box}}2\right)
&
\frac a{l\sqrt{\Box}}  \displaystyle{\frac 1{J_1[l\sqrt{\Box}/a]}} \\[5mm]
\frac a{l\sqrt{\Box}} \displaystyle{\frac 1{J_1[l\sqrt{\Box}/a]}} &
- \frac a{l\sqrt{\Box}} \displaystyle{\frac{J_2[l\sqrt{\Box}/a]}{J_1[l\sqrt{\Box}/a]}}
\end{array}
\right] \, . \label{Gsmalla}
\end{equation}
For convenience we use $\Box$ as the fundamental variable of the expansion and
not $\sqrt\Box$.  The Mittag-Leffler expansion for an element
$G_{\times\times}$ of the Green-function matrix reads
\begin{equation}
G_{\times\times}(\Box)
= \sum^\infty_{i=0} \frac{\Box \text{Res}[G_{\times\times}
(\Box = m_i^2)]}{m_i^2[\Box -  m_i^2]},
\end{equation}
where the first pole is at $\Box = 0$, i.\,e. at $m_0 = 0$.
One can write
\begin{equation}
G_{\times\times}(\Box)
= \sum^\infty_{i=0} \frac{\text{Res}[G_{\times\times}(\Box = m_i^2)]}
{\Box - m_i^2}
+ \sum^\infty_{i=0} \frac{\text{Res}[G_{\times\times}(\Box = m_i^2)]}{m_i^2}. 
\end{equation}
As the second sum contributes only a constant, we can drop this part and
obtain our final result for the elements of the Green function,
\begin{equation}
G_{\times\times}(\Box)
= \sum^\infty_{i=0} \frac{\text{Res}[G_{\times\times}(\Box = m_i^2)]}
{\Box - m_i^2}.
\label{MLGform}
\end{equation}
In this way, each nonlocal element of the nonlocal Green function ${\bf
  G}(\Box)$ can be represented as an infinite sum of scalar propagators with
different masses. The mass-squares are given by the positions of the poles of
${\bf G}(\Box)$.  From the representation of the Green function
(\ref{Gsmalla}), we immediately infer that all elements of ${\bf G}(\Box)$ have
poles at the same values of $\Box$.  This is a direct consequence of the
common prefactor displayed in
(\ref{4.10}) which contains all poles of ${\bf G}(\Box)$. One pole is located
at $\Box = 0$, which is also an exact pole of (\ref{4.10}). An infinite
sequence of poles is located at the roots of the Bessel function $J_1$.
Denoting the argument at the $i$th zero of $J_1$ by $j_i$, we find that these
poles are located at
\begin{equation}
\Box = (j_i a/l)^2 \equiv m_i^2,
\label{poles}
\end{equation}
where the first few values of $j_i$ are given by
\begin{equation}
j_1 \approx 3.832 \, , \quad j_2 \approx 7.016 \, , \quad
j_3 \approx 10.173 \, .
\label{values}
\end{equation}
Whereas the first term of the sum (\ref{MLGform}) from the pole at 
$\Box = 0$ will
provide us with the effective four-dimensional massless graviton, the infinite
series of poles at $m_i$ is responsible for the generation of the KK tower of
massive gravitons. The matrix of residues of {\bf G} at the pole $\Box = 0$ is
given by
\begin{equation}
\text{\bf Res}[{\bf G}(\Box = 0)]
= \frac{2}{l^2(1-a^2)}\left[\begin{array}{cc}
1   & a^2 \\
a^2 & a^4
\end{array}\right]\, ,
\end{equation}
and the residues for the infinite sequence of poles at $\Box = m_i^2$
are found to be
\begin{equation}
\text{\bf Res}[{\bf G}(\Box=m_i^2)]
= \frac{2a^2}{l^2}\left[\begin{array}{cc}
1 / (J_2[lm_i/a])^2 &- 1/ J_2[lm_i/a]\\
- 1/ J_2[lm_i/a] & 1
\end{array}\right].
\end{equation}
The residues can be factorized as
\begin{equation}
\text{\bf Res}[{\bf G}(\Box = m_i)]
= {\bf v_i v_i^T} \, ,\label{ResMLG}
\end{equation}
with
\begin{equation}
    {\bf v}_0= \frac{\sqrt 2}{l\,\sqrt{1-a^2}} 
    \left[\begin{array}{c}  
    \,1 \\  
    \,\,a^2\,  
    \end{array}\right]\, , \quad
    {\bf v}_i= \frac{{\sqrt 2}a}l 
    \left[\begin{array}{c}  
    1/J_2[lm_i/a] \\  
    -1\,  
    \end{array}\right]\ \, , \quad i \geq 1 . \label{vofG}
\end{equation}
This property reflects the fact that the residues of the poles of the Green
function are projectors on the corresponding propagating modes of the theory.
These energy eigenmodes will be recovered from the action in the next section.
First, however, we should display explicitly our newly acquired
representation of the Green function:
\begin{multline}
{\bf G}(\Box) =
\frac 2{l^2\Box (1-a^2)}
\left[ \begin{array}{cc}
1 & a^2\\
a^2 & a^4
\end{array} \right]\\
+ \sum^\infty_{i=1}
\frac {2 a^2}{l^2 (\Box - m_i^2)}
\left[ \begin{array}{cc}
1/(J_2[lm_i/a])^2 & - 1/(J_2[lm_i/a]) \\
- 1/(J_2[lm_i/a]) & 1
\end{array} \right] \, .
\label{MLG}
\end{multline}

%
%

\section{The spectrum and the eigenmodes of the effective action \label{actionsec}}

The eigenmodes of the equations of motion
can be recovered from the action (\ref{4.24}) and its kernel ${\bf F}(\Box)$
(\ref{4.11}). We will find that the situation has many
similarities with resonance theory, and our discussion will
parallel many of the considerations of \cite{More} from which we will also
adapt our nomenclature of eigenstates.

In the low-energy limit, $l \sqrt{\Box} \ll 1$ but $l \sqrt{\Box}/a$
arbitrary, the kernel of the action ${\bf F}(\Box)$ can be approximated by
    \begin{equation}  
    {\bf F}(\Box) \approx \frac{l^2\Box}{2} 
    \left[\begin{array}{cr}  
    1   
    &1/J_2[l\sqrt\Box/a]\\[3mm]   
    1/J_2[l\sqrt\Box/a]&  
    \displaystyle{-\frac 2{l\sqrt{\Box} a}  
    \frac{J_1[l\sqrt\Box/a]}{J_2[l\sqrt\Box/a]}}\,  
    \end{array}\right].      \label{7.4.3}  
    \end{equation} 
  
    The typical way to extract the particle content from an action with a
    matrix-valued kernel is its diagonalization in terms of normal modes.
    However, as found by the application of the Mittag-Leffler expansion of 
    the Green function in
    the last section, the number of propagating modes enormously exceeds the
    number of entries in the $2\times2$-matrix ${\bf F}(\Box)$ and 
    thus the modes
    do not diagonalize the quadratic action (\ref{4.24}) in the
    usual sense. This behavior could also have been anticipated from the
    nonlocality of ${\bf F(\Box)}$.  The propagating modes are the zero modes
    of ${\bf F}(\Box)$ which solve the matrix-valued nonlocal equation
\begin{equation}  
{\bf F}(\Box){\bf h}_i(x)= {\bf 0} \, .
\label{0vec}
\end{equation}
These modes are conveniently split into a scalar part
$h_i(x)$, depending on the four-dimensional spacetime coordinates, 
and a spacetime-independent (isotopic) vector part ${\bf v}_i$,
\begin{equation}
{\bf h}_i(x)=h_i(x){\bf  v}_i \, . \label{siegertmode}
\end{equation}
The eigenvector equation (\ref{0vec}) is accompanied by its consistency
condition,
\begin{equation} 
\det{\bf F}(\Box)=0, \label{consist}
\end{equation}
which corresponds to picking the poles of the Green function ${\bf G}(\Box)$.
The condition (\ref{consist})
yields the mass spectrum of the theory given by 
the roots of this equation, i.\,e.\ $\Box=m_i^2$. Hence the $h_i(x)$ above 
are Klein-Gordon modes and the isotopic 
vectors of the propagating modes ${\bf v}_i$ are zero eigenvectors 
of ${\bf F}(m_i^2)$,
\begin{flalign}
&(\Box-m_i^2)\,h_i(x)=0,\\
&{\bf F}(m_i^2){\bf v}_i=0.
\end{flalign}
Thus we obtain the Kaluza-Klein spectrum which contains the massless mode
$i=0$, $m_0=0$, and a tower of massive modes. In the low-energy
approximation of ${\bf F}(\Box)$, see Eq.\ (\ref{7.4.3}), its masses $m_i=a
j_i/l$ are given by the roots of the first-order Bessel function,
$J_1(j_i)=0$. The isotopic structure of their ${\bf v}_i$ is given by the
vectors (\ref{vofG}) which were found to factorize the residues of the
Green function.\footnote{ The vector ${\bf v}_0$ is here actually given by
  ${\bf v}_0 = \sqrt 2/l [1 \,\, a^2]^T$, i.\,e. the leading order in $a$ of
  the exact ${\bf v}_0$ as given by (\ref{vofG}), because here the ${\bf
    v}_i$'s are obtained for the approximation (\ref{7.4.3}) of ${\bf F}$.}
In view of standard arguments of gauge invariance, the massless graviton $h_0$
has two dynamical degrees of freedom, while the massive tensor field $h_M$
posesses all five polarizations of a generic transverse-traceless tensor
field.

The action (\ref{4.24}) is not, however, diagonalizable in the basis of these 
states because under the decomposition ${\bf h}(x)=\sum_{i=0}h_i(x){\bf v}_i$ 
(with off-shell coefficients $h_i(x)$) the cross terms intertwining 
different $i$'s are non-vanishing, 
\begin{equation}
{\bf v}_i^T {\bf F}(\Box) {\bf v}_j\neq 0. 
\end{equation} 
A crucial observation is, however, that the diagonal and non-diagonal 
terms of this expansion are linear and bilinear, respectively, in on-shell  
operators $\Box-m_i^2$,  
\begin{align} 
&{\bf v}^T_i {\bf F}(\Box) {\bf v}_i  
= \Big({\bf v}^T_i \frac{d{\bf F}(\Box)}{d\Box} 
{\bf v}_i\Big)_{\Box = m_i^2} (\Box - m_i^2),\label{99}\\ 
&{\bf v}^T_i {\bf F}(\Box) {\bf v}_j  
= M_{ij}(\Box)(\Box - m_i^2)(\Box - m_j^2),\,\,i\neq j, \label{100} 
\end{align} 
where higher powers of $(\Box-m_i^2)$ have been dropped in (\ref{99}), and
$M_{ij}(\Box)$ is non-vanishing at both $\Box=m_i^2$ and $\Box=m_j^2$.
Therefore, the non-diagonal terms of the action do not contribute to the
residues of the Green function ${\bf G}(\Box)$ of ${\bf F}(\Box)$.  The
normalization of ${\bf v}_i$, as given by (\ref{vofG}), automatically yields 
in (\ref{99}) a
unit coefficient of $(\Box-m_i^2)$.

Thus we have found the particle spectrum of the action in terms of modes which
turn out to be non-orthogonal off-shell but become orthogonal if one considers
the action on the mass-shell. In the theory of atomic and nuclear resonances,
such non-orthogonal energy eigenstates are called {\em Siegert states} 
after their introduction in \cite{Siegert} (cf.\ \cite{More}).

This particle spectrum can also be recovered from the two {\em orthogonal}
eigenstates of ${\bf F}(\Box)$. These states which will be energy dependent,
i.\,e.\ {\em not} mass eigenstates, have also been extensively used in
resonance theory and are named {\em Kapur-Peierls states} 
according to their first use in \cite{Kapur,Peierls} (cf.\ also \cite{More}).

In resonance theory, Kapur-Peierls states ${\bf b}_i(\Box)$ diagonalize the
{\em matrix} Hamiltonian of the system (the analogue of our ${\bf F}(\Box)$).
For our $2 \times 2$ matrix-valued kernel ${\bf F}(\Box)$, this property reads
\begin{flalign}
&{\bf F}(\Box) {\bf b}_s(\Box) = \lambda_s(\Box) {\bf b}_s(\Box)\,, 
\quad s=1,2, \label{eigeneq} \\ 
&{\bf b}^T_s(\Box){\bf b}_{s'}(\Box)=\delta_{ss'},
\label{ortho}
\end{flalign}
where the second equation describes the orthonormality of the Kapur-Peierls
states.

In the limit $l\sqrt\Box \ll 1$, the energy-dependent eigenvalues of ${\bf
  F}(\Box)$ [as determined by Eq.\ (\ref{4.11})] are found to be
\begin{equation}
\lambda_1(\Box) = \frac{l^2 \Box}2 + O(a) \, , \quad
\lambda_2(\Box) = \frac{l \sqrt\Box}a \frac{J_1[lm_i/a]}{J_2[lm_i/a]}
+O(a)\, . \label{eigenval}
\end{equation}
The corresponding eigenvectors (normalized to unity to leading order in $a$) are
\begin{align}
{\bf b}_1(\Box) = \left[\begin{array}{c}1 \\[1mm] 
\displaystyle{\frac{a\, l \sqrt \Box}{2J_1[lm_i/a]}}
\end{array}\right] \, , \quad
{\bf b}_2(\Box) = \frac 1{\sqrt{1-Z^2}} 
\left[\begin{array}{c}1 \\ - Z
\end{array}\right] ,
\end{align}
where we have introduced 
\begin{equation}
Z \equiv  \left| \frac{l\sqrt\Box}a \left[
\frac 2{l^2\Box} +\frac 12 \right] J_1[lm_i/a] - J_2[lm_i/a] \right|.
\end{equation}
The eigenvalue $\lambda_1$, given in (\ref{eigenval}), has only one root at
the value $\Box = 0$, whereas the eigenvalue $\lambda_2$ has infinitely many
zeros at the zeros of the Bessel function $J_1$, i.\,e.\ at $\Box = m_i^2$.
Approximating $\lambda_2(\Box)$ around its zeros by the first term of its
Taylor expansion, we find for the eigenvalues
\begin{equation}
\lambda_1(\Box) = \frac{l^2 \Box}2
\, , \quad
\lambda_2(\Box) = \frac{l^2}{2 a^2} (\Box -m_i^2) \, .
\end{equation}
At $\Box = 0$ the eigenvector ${\bf b}_1$ takes the value
\begin{equation}
{\bf b}_1(0) = \left[\begin{array}{c} 1 \\ a^2 \end{array}\right] \, ,
\end{equation}
and  at energies $\Box = m_i^2$ the eigenvector ${\bf b}_2$ becomes
\begin{equation}
{\bf b}_2(m_i^2) = \left[\begin{array}{c} 1 \\ - J_2[lm_i/a]\end{array}\right].
\end{equation}
As expected, the mass levels of the Siegert states equal the zeros of the
Kapur-Peierls eigenvalues. The massless graviton Siegert state corresponds to
the single root of the eigenvalue $\lambda_1(\Box)$, and the tower of massive
Siegert states arises from the infinitely many zeros of the eigenvalue
$\lambda_2(\Box)$.  It is easy to find the relation between the Siegert
eigenvectors and the Kapur-Peierls eigenvectors by taking from Eq.\ 
(\ref{eigeneq}) that the vectors ${\bf b}_s(\Box)$ are eigenvectors also of
${\bf F}^{-1}(\Box) = {\bf G}(\Box)$ with eigenvalues $1/\lambda_s(\Box)$.
Using the completeness and orthonormality of the ${\bf b}_s(\Box)$'s, we obtain
a representation of ${\bf G}(\Box)$,
\begin{equation}
{\bf G}(\Box) =
\sum_{s=1,2} \frac{{\bf b}_s(\Box){\bf b}^T_s(\Box)}{\lambda_s(\Box)}\, .
\end{equation}
As the poles of the
Green function lie at $\lambda_s(\Box) = 0$ we find for the residues of 
${\bf G}(\Box)$
\begin{equation}
\lim_{\Box \to m_i^2} (\Box -m_i^2){\bf G}(\Box) =
\left. \frac{{\bf b}_s(\Box){\bf b}^T_s(\Box)}
{d\lambda_s(\Box)/d\Box }\right|_{\Box = m_i^2}\, ,
\end{equation}
where $s=1$ for $m_0 \equiv 0$ and $s=2$ for $m_i$ with $i \geq 1$.
Comparing this expression for the residues with (\ref{ResMLG}), we obtain
\begin{equation}
{\bf v}_0 = \left. \left( \frac{d \lambda_1(\Box)}{d\Box} \right)^{-1/2} \,
            {\bf b}_1(\Box) \right|_{\Box = 0} \, ,\quad 
{\bf v}_i =  \left. \left( \frac{d \lambda_1(\Box)}{d\Box}  \right)^{-1/2} \,
            {\bf b}_2(\Box) \right|_{\Box = m_i^2} \, , \quad i\geq 1.
\label{eigenrel}
\end{equation}
 
The decomposition of the action into Siegert eigenmodes (\ref{siegertmode})
provides us with the conventional  
particle interpretation of the propagating modes of the nonlocal operator 
${\bf F}(\Box)$. It clarifies their role in its Green function (\ref{MLG}) 
that mediates the gravitational effect of matter sources. Amended by the  
matter action on the branes, the effective action of the  
graviton sector reads 
\begin{align} 
  S\,[\,h^\pm_{\mu\nu}\,]  
    =\int  
    d^4x\,\left(\frac1{32\pi G_4}  
    {\bf h}^T  
    \,\frac{{\bf  F}(\Box)}{l^2}\,  
    {\bf h}+  \frac12{\bf h}^T {\bf T} \right), 
\end{align} 
where ${\bf T}$ is the column vector of the transverse-traceless part the
stress-energy tensors on the branes
and ${\bf h}$ is now given by the sum of Siegert modes (\ref{siegertmode}),
\begin{equation}
{\bf h} = \sum_{i=0} h_i(x) {\bf v}_i\, .
\label{hcomp}
\end{equation}
Varying this action with respect to each
Siegert mode ${\bf h}_i$ and decomposing the results to recover the equations
of motion for ${\bf h}$, we obtain the linearized equations of motion. Their
solution
\begin{equation}   
{\bf h} =-8\pi G_4 l^2{\bf G}_\text{ret} {\bf T}
\end{equation}
is expressed 
in terms of the retarded version of the Green function (\ref{MLG}).   
We note the explicit expressions for $h^+$ and $h^-$ obtainable from 
(\ref{eigenrel}) and (\ref{hcomp}):
\begin{flalign}
h^+(x) &= \frac{\sqrt 2}{l} \left( 
\frac 1{\sqrt{1-a^2}} h_0(x) + a \sum_{i_1}\frac 1{J_2[lm_i/a]}h_i(x)
\right),
\label{mix+} \\
h^-(x) &= \frac{\sqrt 2}{l} \left( 
\frac {a^2}{\sqrt{1-a^2}} h_0(x) - a \sum_{i_1} h_i(x)
\right).
\label{mix-}
\end{flalign}
It is particularly interesting to study the asymptotic behavior of the
expressions (\ref{mix+}) and (\ref{mix-}).  In the long-distance limit
$a \rightarrow 0$, we have
\begin{align}
h^+ &\sim h_0 + O(a)\, ,\\
h^- &\sim a^2 h_0 - a \sum\nolimits_{i\geq 1} h_i \, . \label{mixing5}
\end{align}
Thus, in this limit, the $h^+$-mode practically coincides with the massless
graviton $h_0$, and the mixing with the other modes is suppressed by the
factor $a$. On the other hand, the $\Sigma_-$-brane mode $h^-$ is almost
exclusively composed from the massive modes, and mixing with the massless
mode is suppressed, whereas the mixing between the massive modes can be called
maximal, i.\,e.\ $h^-$ is proportional to the sum of the massive modes.
Moreover, the contribution of both the modes $h_i$ and $h_0$ to the $h^-$-mode
in this regime is suppressed by the common factor $a$.  
There is also a regime of maximal mixing between {\em all} modes for the
metric perturbation $h^-$ on the $\Sigma_-$-brane at $a = 1/\sqrt 2$ where
\begin{equation}
h^- \sim h_0 + \sum\nolimits_{i\geq 1} h_i \, .
\label{maxmix-}
\end{equation}

\section{The graviton effective action in the diagonalization approximation
\label{diag}} 

In the last section we have found that the Siegert state representing the 
massless
graviton corresponds to the zero of the eigenvalue $\lambda_1$ of the kernel
of the action ${\bf F}(\Box)$, whereas the Siegert states corresponding to all
of the KK modes stem from the zeros of the second eigenvalue $\lambda_2$.
This opens the possibility of a low-energy limit in which the Siegert and
Kapur-Peierls states coincide.  In the approximation
\begin{equation}
\sqrt{\Box} \ll 1 \, , \quad l/a\sqrt{\Box} \lesssim 4 \ ,\label{lc}
\end{equation}
one can use the small-argument expansion for all Bessel and Neumann functions
appearing in ${\bf F}(\Box)$.  If we truncate this expansion at $O(\Box)$, we
obtain an expression for the kernel ${\bf F}(\Box)$ which is entirely local.
The choice of the second part of the low-energy condition (\ref{lc}) guarantees
that the approximation is still valid slightly above the mass of the first
KK mode,
\begin{equation}
m_1 = j_1 a/l  \quad \Leftrightarrow \quad l m_1 /a \approx 3.83  
\end{equation}
(cf.\ \ref{values}).
In fact, at least for Bessel functions of the second kind, the small-argument
expansion is considered as the method of choice in numerical calculations for
arguments $\lesssim 4$, see \cite{Arfken}. However, we should be prepared for some
deviation of the expansion-result for the first KK mode from that of the
previous sections due to the low order of truncation of the expansion. Taking
this into account, the result will turn out to be surprisingly accurate.

In the low-energy approximation (\ref{lc}), the kernel ${\bf F}/l^2$ can be
represented in the form \cite{we}:
\begin{align}
    \frac{\bf F(\Box)}{l^2}&=
    -{\bf M}+{\bf D}\Box+
    O(\Box^2),    \label{oper-gr}\\
    {\bf M}&=\frac1{l^2}\,\frac8{1-a^4}\,
    \left[\begin{array}{cc}
    \,a^4\, & \,-a^2 \\
    -a^2 & \,1 \end{array}\right],      \label{oper-gr1}\\
    {\bf D}&=\frac{1-a^2}{3(1+a^2)^2}
    \left[\begin{array}{cc}
    \,a^2+3\, & 2 \\
    2 & \,3+a^{-2}
    \end{array}\right].                 \label{oper-gr2}
\end{align}
    
The operator ${\bf D}$ represents the kinetic part of the action (\ref{4.24}),
while the operator ${\bf M}$ plays the role of the mass term.  Calculating the
trace and the determinant of the matrix ${\bf D}$, it is easy to see that it
is positive definite for any value of the parameter $a$.  The determinant of
the matrix ${\bf M}$ is equal to zero, therefore ${\bf M}$ is degenerate.
Using the positive-definiteness of the matrix ${\bf D}$, one can diagonalize
both the matrices ${\bf D}$ and ${\bf M}$ simultaneously.

Our purpose here is not simply to diagonalize these matrices, but also to
make a canonical normalization of the kinetic terms. With this choice, the
action for gravity eigenmodes with fixed masses, representing some mixtures of
the functions $h^+$ and $h^-$, will have the form
\begin{equation}
S_{\rm graviton}
[h_0,h_M] = \frac{1}{16\pi G_4}\int d^4x
\left(\frac14 h_0\Box h_0 + \frac14 h_M(\Box - M^2)h_M\right),
\label{action-new}
\end{equation}
where $M$ is the mass of the massive gravitational mode. 

The details of the procedure of diagonalization are presented in the Appendix.
Here we shall give the final results.  The expression for the mass of the
massive graviton mode $h_M$ is
\begin{equation}
M^2 = \frac{24 a^2 (1+a^2)}{l^2 (1-a^2)^2}. \label{Mass0}
\end{equation}
At small
values of the parameter $a$, i.\,e.\ at large distances between the branes,
the squared mass behaves as
\begin{equation}
M^2 \sim \frac{24a^2}{l^2} \label{Mass1}
\end{equation}
and tends to zero together with $a^2$.  This result is to be compared with the more accurate
expression for the first KK mass in relation (\ref{poles}). The
large discrepancy is easily explained from the fact that taking the
limit $a\to 0$ is strictly speaking not legitimate in our approximation
because the second condition of (\ref{lc}) would not be fulfilled anymore.

The relations describing the transition from the old graviton
variables $h^+$ and $h^-$ to the modes $h_M$ and $h_0$ have the following form,
\begin{align}
h_M &= \frac{\sqrt{1-a^2}}{\sqrt{3}(1+a^2)}\left(ah^+ -
\frac{h^-}{a}\right),\label{mixing30} \\
h_0 &= \frac{\sqrt{1-a^2}}{1+a^2}(h^+ + h^-). \label{mixing3}
\end{align}
The inverse transformations are
\begin{align}
h^+ &= \frac{1}{\sqrt{1-a^2}}(h_0 + \sqrt{3}a h_M),\label{mixing40}\\
h^- &= \frac{a}{\sqrt{1-a^2}}(ah_0 -\sqrt{3}h_M).
\label{mixing4}
\end{align}
The corresponding expressions for ${\bf v}_0$ and ${\bf v}_M$ are
\begin{equation}
{\bf v}_0 = \frac{\sqrt{2}}{l\sqrt{1-a^2}}
\left[ \begin{array}{c}1\\a^2\end{array} \right] \, , \quad
{\bf v}_M =\frac{\sqrt{6}a}{l\sqrt{1-a^2}}
\left[ \begin{array}{c}1\\ -1\end{array} \right] \, ,
\end{equation}
which should be compared with the expressions for ${\bf v}_0$ and ${\bf v}_1$
in (\ref{vofG}). As might have been anticipated from the fact that the
residues depend only on the $(-1)$th term of the Laurent expansion of the
function considered, we find exact agreement for ${\bf v}_0$. 

Amending the action (\ref{action-new}) 
by the matter action (\ref{matteraction}),
we can obtain the equations of motion for the metric perturbations $h^+$
and $h^-$ on the brane by varying the total action with respect to $h_0$
and $h_M$ and then combining their equations of motion according to the
transformations (\ref{mixing40}) and (\ref{mixing4}),
\begin{multline}
\left[\begin{array}{c}h^+ \\ h^- \end{array}\right]
= - 16 \pi G_4 \left( \left. \frac 1{\Box}\right|_\text{ret} \frac 1{1-a^2}
\left[\begin{array}{cc}1&a^2\\a^2&a^4\end{array}\right] \right. \\ \left.
+ \left. \frac 1{\Box -M^2}\right|_\text{ret} \frac{3a^2}{1-a^2}
\left[ \begin{array}{cc}1&-1\\-1&1\end{array} \right]
\right) \, \left[\begin{array}{c}T^+ \\ T^- \end{array}\right].
\end{multline}
In the language of the Siegert and Kapur-Peierls eigenmodes of
Sec.~\ref{actionsec}, we have chosen for the diagonalization an approximation
in which the Siegert eigenmodes become orthogonal and the
Kapur-Peierls eigenvalues and eigenstates energy-{\em independent}.
Thus, in the approximation (\ref{lc}), the Siegert and Kapur-Peierls
descriptions coincide.
The first Kapur-Peierls state coincides with the Siegert state of
the massless graviton, while the massive Siegert states become a kind of
``collective'' massive state which coincides with the second Kapur-Peierls 
mode.

%
%

\section{Graviton oscillations \label{phen}}

\subsection{Gravitational waves on the $\Sigma_+$-brane \label{GW+}}

After the extensive theoretical considerations of Secs.\ \ref{actionsec} and
\ref{diag}, we will now return to the equations of motion (\ref{eom}) and
study their phenomenological properties. We will proceed to show how
graviton oscillations arise naturally in the propagations of gravitational
waves in this model. For studying the properties of graviton oscillations, it
is convenient to abandon the condensed matrix notation of Eq.~(\ref{eom}) and
to study the behavior of metric perturbations on the $\Sigma_+$-brane and
$\Sigma_-$-brane separately, starting with the $\Sigma_+$-brane.

Using the spectral representation (\ref{MLG}), we find for $h^+$,  
      \begin{equation} 
       h^+= -\left.\frac{16\pi G_4}{\Box} 
       \right|_{\rm ret} 
(T^++a^2 T^-) 
        - \sum^\infty_{i=1} \left.\frac{16\pi G_4}{\Box-m_i^2} 
        \right|_{\rm ret} 
\left(\frac{a^2}{\mathcal{J}_2^2}T^+- 
       \frac{a^2}{\mathcal{J}_2}T^-\right) \, ,
\end{equation} 
where we have introduced $\mathcal{J}_2 \equiv J_2[lm_1/a] \approx 0.403$. 
We consider astrophysical sources of equal  
intensity at ${\mathbf x}=0$ on both branes with a harmonic time dependence
\begin{equation}  
T^\pm(t,{\mathbf x})=\mu e^{-i\omega t}\delta({\mathbf x})\, , \label{source}
\end{equation}
where $\mu$ is the mass quadrupole moment of the source.
The sources with an oscillating quadrupole moment will generate gravitational
waves.  At a distance $r$ from the source, the waves on each brane are given by
a mixture of massless and massive spherical waves.  On the $\Sigma_+$-brane, 
this superposition is given by the sum of the contributions
from the sources on the $\Sigma_+$- and $\Sigma_-$-brane, respectively,
\begin{align}
 h^+[\,T^+] & = A\, e^{-i\omega t} 
 \left(e^{i \omega r} +  \sum_{i=1} \frac{a^2}{(J_2[lm_i/a])^2}  
 e^{i \sqrt{ \omega^2-m_i^2}r} \right) \, , \label{h+of+} \\  
 h^+[\,T^-] & = A a^2\, e^{-i\omega t}  
 \left( e^{i \omega r} -  \sum_{i=1} \frac{1}{J_2[lm_i/a]}  
 e^{i \sqrt{ \omega^2-m_i^2}r} \right) \, , \label{h+of-}
\end{align}
where
$A=4\,G_4\,\mu/r$ is the amplitude of the massless mode generated on the
$\Sigma_+$-brane.
If the frequency $\omega$ with which the source is oscillating is smaller than
the mass $m_i$ of a KK mode, $\omega < m_i$, then the exponent of the
corresponding mode will be real and negative, instead of imaginary, and the mode
will be decaying instead of oscillating. In effect all modes with masses above
the frequency of the source will have died off after propagating for a short
distance and only those modes with a mass below the frequency of the source
will contribute to the gravitational wave in the far field.
Thus, as long as we are interested in GW's, we can content ourselves with
including only the terms with $m_i < \omega$ into the sums of Eqs.\ (\ref{h+of+})
and (\ref{h+of-}) (cf.\ \cite{GRS}).

In order to have a simple situation, we consider a source with a frequency
above the mass threshold of the first massive mode but below the threshold of
the second KK mode, i.\,e.\ $m_1 < \omega < m_2$. Then only the
massless and the first massive mode are excited and produce long-range
gravitational waves.  At a distance $r$ from the source, the waves on each brane
are given by a mixture of massless and massive spherical waves. 
On the 
$\Sigma_+$-brane this superposition is given by the sum of the  
contributions  
 \begin{flalign} 
 h^+[\,T^+] & = A\, e^{-i\omega t} 
 \left(e^{i k r} +  \frac{a^2}{\mathcal{J}_2^2}  
 e^{i \sqrt{ k^2-m_1^2}r} 
                              \right) \, ,
\\  
 h^+[\,T^-] & = A a^2\, e^{-i\omega t}  
 \left( e^{ik r} - \frac{1}{\mathcal{J}_2}  
 e^{i \sqrt{k^2-m_1^2}r} \right) \, ,          
 \end{flalign} 
from sources on the $\Sigma_+$- and $\Sigma_-$-brane, respectively. 

\subsection{Quantum  oscillations --- an analogy}\label{quantum}

The above results have interesting physical consequences. We have seen that
the \emph{observable} transverse-traceless metric on each brane is indeed a
linear combination of different gravitational modes with fixed masses.  One of
these modes appears to be massless, while the others are massive with their
masses depending on the distance between the branes. 

This situation is quite analogous to the well-known mixing arising in the
context of kaons \cite{kaon,lipkin} or other bosons \cite{boson} or with neutrino
oscillations \cite{neutrino}.  As a matter of fact, the phenomenon which we
encounter here is not quantum oscillations but \emph{classical} ones.
Nevertheless, the interference effects between the different gravitational
modes, arising in the process of their propagation in space (on the brane),
are analogous to those of quantum oscillations. This is quite natural because
the oscillations of quantum particles are totally conditioned by their wave
nature (see, for example, Ref. \cite{lipkin}), and both effects are caused by
the superposition of modes with different dispersion relations.

However, there is one important difference: 
it does not matter for quantum oscillations if one considers the superposition
to consist of two components of different energies but with the same momentum
(as is usually assumed) or if the components are supposed to have the same
energy but different momenta \cite{MoPa}. For a mixture of classical
gravitational waves, however, it is natural to assume that they have been
generated by the same source oscillating with a frequency $\omega$. Therefore
one should consider oscillations between modes of the same frequency $\omega$
which have different wave-numbers $k$, and not vice versa.

Let us recall briefly some basic formulas from the theory of quantum
oscillations. Suppose that a certain particle can exist in two different
observable states $|\psi_1\rangle$ and $|\psi_2\rangle$, which correspond to
a quantity conserved in some interaction but which do not have
fixed masses. For example, kaons can be created or detected in states with
fixed strangeness, which are not eigenstates of their Hamiltonian, and
according to the theory of neutrino oscillations, which is now supported by
good experimental evidence \cite{neu-exp}, the states of the electron, muon
and tau neutrinos are also mixtures of different Hamiltonian (i.\,e.\ mass)
eigenstates. Generally, this situation can be described for the two-state case
by
\begin{align}
|\psi_1\rangle &= \cos\theta|\psi_M\rangle +
\sin\theta|\psi_m\rangle, \nonumber \\
|\psi_2\rangle &= -\sin\theta|\psi_M\rangle +
\cos\theta|\psi_m\rangle, \label{osc}
\end{align}
where the states $|\psi_M\rangle$ and $|\psi_m\rangle$ are states
with masses $M$ and $m$, respectively, and the angle $\theta$ parameterizes
the mixing between the states. If at the initial position of an
evolution at $x=0$
somebody observes, say,  a particle in a state
\begin{equation}
|\psi(0)\rangle = |\psi_1\rangle, \label{initial}
\end{equation}
propagating in space with the energy $\omega$, then at the position $x$
the state (\ref{initial}) will become equal to
\begin{align}
|\psi(x)\rangle &= \cos\theta\, e^{-ik_Mx}|\psi_M\rangle +
\sin\theta\, e^{-ik_mx}|\psi_m\rangle \nonumber \\
\phantom{|\psi(x)\rangle} &= \cos\theta\, e^{-i\sqrt{k_0^2+M^2}x}|\psi_M\rangle +
\sin\theta\, e^{-i\sqrt{k_0^2+m^2}x}|\psi_m\rangle, \label{final}
\end{align}
where $k_0 = \omega$.
Now we are in a position to calculate the probability of observing
the particle in the state $|\psi_1\rangle$ at the position $x$ which
is given by the formula
\begin{equation}
P_1(x) = |\langle \psi_1|\psi(x)\rangle|^2 =
1 - \sin^2 (2\theta) \, \sin^2 \frac{(k_M - k_m) x}{2}\, . \label{probab}
\end{equation}
Correspondingly, the probability of finding the particle in the state
$|\psi_2\rangle$ is
\begin{equation}
P_2(x) = |\langle \psi_2|\psi(x)\rangle|^2 =
\sin^2 (2\theta)\, \sin^2 \frac{(k_M - k_m) x}{2}. \label{probab1}
\end{equation}

It is easy to see that the oscillations of the probabilities $P_1$ and
$P_2$ are strongest when $\sin^2 (2\theta) = 1$, i.\,e.\ when
\begin{equation}
\theta = \pm \frac{\pi}{4}. \label{maxosc}
\end{equation}
This situation is called the case of maximal mixing between
oscillating particles. Looking at the expressions for the probabilities
(\ref{probab}) and (\ref{probab1}), one can see that they represent
periodic functions with a period
\begin{equation}
L=\frac{2\pi}{\sqrt{k_0^2+M^2}-\sqrt{k_0^2+m^2}}. \label{neu-length}
\end{equation}
This is the oscillation length of a quantum state with the energy 
$\omega = k_0$.

In the case of maximal mixing (\ref{maxosc}), detecting the particle
$|\psi_1\rangle$ at the distance $L/2$, one finds with
probability $P_2(L/2)=1$ the particle in the state $|\psi_2\rangle$.  In
other words, the particle in the state $|\psi_1\rangle$ is totally transformed
into the particle in the state $|\psi_2\rangle$. The time-dependence of the 
oscillation can be inferred from this expression by considering semi-classically the propagation speed of the corresponding particles.

The description of quantum oscillations presented above
coincides with describing classical oscillations of superpositions of
waves with different dispersion relations. In this case, instead of
the probability of detecting a particle of a certain type, one can calculate
the amplitude of the wave. 

\subsection{Gravitational-wave oscillations on the $\Sigma_+$-brane \label{RIGO+}}

For the superpositions (\ref{h+of+}) and (\ref{h+of-}),
the amplitudes detected by a gravitational-wave interferometer are given by
the absolute values of the waves,
\begin{align} 
\big|h^+[\,T^+]\big| &= \mathcal{A}^+ 
\left[ 1- \frac{4a^2 \mathcal{J}^2_2}{(\mathcal{J}^2_2+a^2)^2}  
\sin^2 \left( \frac{\pi r}{L}\right) 
\right]^{1/2}, 
\label{absh+of+} 
\\ 
\big|h^+[\,T^-]\big| &=  
\mathcal{A}^- 
\left[ 1 + \frac{4\mathcal{J}_2}{(\mathcal{J}_2-1)^2}  
\sin^2 \left( \frac{\pi r}{L}\right) 
\right]^{1/2}. 
\label{absh+of-} 
\end{align} 
Here, $L$ is the oscillation length of the amplitude modulation of the 
gravitational wave (GW),
\begin{equation} 
L= \frac{2\pi}{\omega - \sqrt{\omega ^2-m_1^2}} \, .\label{length} 
\end{equation} 
This corresponds to
Eq.\ (\ref{neu-length}) with one mode being massless.
 The pre-factors
of the amplitudes (\ref{absh+of+}) and (\ref{absh+of-}) are given by
\begin{align} 
\mathcal{A}^+ &= \left(1+ \frac{a^2}{\mathcal{J}^2_2}\right) A \approx A,
\label{prefacA+} \\ 
\mathcal{A}^- &= \left(\frac{1}{\mathcal{J}_2}-1\right) a^2 \,A 
\approx  1.5 \, a^2A, 
\label{prefacA-} 
\end{align} 
where the approximations are valid in the limit $a\ll 1$. We find oscillations
in the amplitudes of the waves from both sources. For a GW generated by $T^+$,
see Eq.~(\ref{absh+of+}), the oscillating part of the amplitude is suppressed by a
factor of $a^2$ compared to the constant part in the limit of
large brane separation, $a \ll 1$. The amplitude of the GW produced by $T^-$,
see Eq.\ (\ref{absh+of-}), is totally oscillating, regardless of the inter-brane
distance.

In the limit of a source-frequency just above the mass-threshold 
$\omega \gtrsim m_1$, the oscillation length (\ref{length}) of these
radion-induced gravitational (``graviton'') 
wave oscillations (RIGO's) tends to
\begin{equation}
L \gtrsim \frac{2\pi}{m_1} \, .
\label{short-length}
\end{equation}
This limit is mainly useful to set a lower bound on the oscillation lengths as
in this limit the propagation speed of the massive mode will tend to
zero. Therefore, in this limit, the massless and massive wavetrains from a
source would soon become spatially separated from each other and, far from the
source, one would detect two separated wavetrains instead of one wavetrain
exhibiting RIGO's. The astrophysically more interesting limit is the case
$\omega \gg m_1$ in which the expression for the oscillation length becomes
\begin{equation}
L = 4 \pi \frac{\omega}{m_1^2} \, .
\label{long-length}
\end{equation}
We can express the minimal oscillation length (\ref{short-length}) of 
RIGO's in terms of the AdS radius $l$ and the scale factor $a$, 
\begin{equation} 
L_{\rm min} = 2 \pi  \, \frac{l}{j_1 a}  \approx 1.6\, \frac{l}a\ , 
\end{equation}
where $j_1$ is the first root of $J_1$ (cf.\ (\ref{poles}) and
(\ref{values})).  The oscillation length is inversely proportional to $a$.
Graviton oscillations become observable when the oscillation length is at
least of the order of the arm length of a GW detector. For the ground-based
interferometric detectors this requirement corresponds to $L \sim 10^3\,{\rm
  m}$.  Combining this with the constraint on the maximal size of the AdS
radius $l$ from sub-millimeter tests of gravity, $l \lesssim 10^{-4}\,{\rm m}$
\cite{Gundlach}, we find an upper limit on the warp factor $a \lesssim
10^{-7}$ for the oscillation length to be detectable.  Inserting this into the
ratio of the amplitudes (\ref{prefacA+}) and (\ref{prefacA-}) we find
\begin{equation} 
\mathcal{A}^- / \mathcal{A}^+  \sim a^2 \lesssim 10^{-14}. 
\label{damping}
\end{equation} 
Therefore, the amplitude of a wave originating from a source on the
(``hidden'') $\Sigma_-$-brane with oscillations which are sufficiently long to
be detectable, is strongly suppressed by a damping factor $a^2$ as compared to a
GW stemming from a source on the $\Sigma_+$-brane itself.  A strongly
oscillating wave has to be generated by a source of 14 orders of magnitude
stronger than that of a weakly oscillating one in order to be of the same
magnitude, which at first sight makes the detection of RIGO's impossible.
We will, however, see in the next subsection that the damping factor could
easily be overcome if the RS model is considered as a model for a putative
M-theory realization of large extra dimensions.

Beforehand we want to summarize that the main differences between
the graviton oscillations considered in this paper and the more
traditional neutrino or kaon oscillations:

First, as is readily inferred from Eq.\ (\ref{hcomp}) or (\ref{mix+}) and
(\ref{mix-}), respectively, the transformation between the gravitational modes
$h^+$ and $h^-$ living on certain branes and the graviton modes $h_0$ and
$h_i$ possessing fixed masses is not an orthogonal (and not a unitary)
transformation. This is connected with the fact that --- in contrast to the
mixing for neutrinos or kaons, where only the mass matrix is non-diagonal if
written in terms of observable states --- in the case of gravity in a
two-brane world the whole kernel of the action is non-diagonal and, moreover,
is diagonalizable only on-shell (cf.\ Sec.\ 4).

The second and, perhaps, more important distinguishing feature of RIGO's is
the dependence of the mixing parameters and the mass of the massive mode on
the parameter $a$, i.\,e.\ their dependence on the radion.  If one considers a
model in which the distance between the branes is time-dependent, the mixing
parameters and all the oscillatory effects become functions of
time depending on the particular features of the model under consideration.

\subsection{High-amplitude RIGO's on the $\Sigma_+$-brane from M-theory}\label{Mtheory}

The bad prospects for the detection of RIGO's found at the end of the last
section may considerably be improved if one considers the RS two-brane model
as a toy model for a yet-to-be-constructed model of the strong-coupling limit
of heterotic M-theory with one large extra dimension. Although such a model is
lacking, we can extrapolate generic features of M-theory compactifications with
small extra dimensions and draw some reasonable conclusions for the hidden
sector of the model. These conclusions suggest the possibility that the hidden
brane could in fact be the source of GW's with very large amplitudes.

Cosmological M-theory models consist of an 11-dimensional spacetime which is
compactified on an $S^1/{\mathbb Z}_2$ orbifold with a stack of D-branes on
each orbifold-fixed plane \cite{HW,WitCoup,LOSW}.\footnote{Cf.\ \cite{Pyr} for
  a non-technical description of the cosmological aspects of this class of
  models.}  Further 6 of the 10 dimensions are compactified on a Calabi-Yau
three-fold, while the other dimensions parallel to the stacks of branes,
including the timelike direction, remain uncompactified. The compactification
scale of the Calabi-Yau manifold is generally assumed to be smaller by one or
two orders of magnitude than the size of the compact 11th dimension. Each
stack of branes hosts one super-multiplet of $E_8$ gauge fields. This
construction can become a suitable model for our universe if one of the set of
$E_8$ fields is broken into the subgroup $E_6$ and further into $SU(3) \times
SU(2) \times U(1)$. These fields are then called the visible sector of the
model and the corresponding stack of branes is identified as our universe,
while the other $E_8$ is called the hidden sector.  In order to obtain a
viable phenomenology it is necessary to break the $E_8$ of the visible sector
by gaugino condensation in the hidden sector and then mediate the influence of
the gaugino condensate to the visible sector by its coupling to moduli fields
on the Calabi-Yau manifold.\footnote{ Cf. \cite{Polchinski}, Chap.\ 18.3, p.\ 
  366--371, for a pedagogical description of this mechanism in the
  weak-coupling limit of $E_8 \times E_8$ heterotic string theory.}  The
volume of the Calabi-Yau space $V$ at the visible sector is assumed to be of
the size of the standard-model grand-unification scale $V^{-1/6} \sim
10^{16}\,{\rm GeV}$. In generic M-theory compactifications, it has been found
necessary in this context to choose the volume of the Calabi-Yau space at the
hidden sector smaller than at the visible sector, which will also lead
to a stronger gauge coupling in the hidden sector than in the visible one
\cite{WitCoup} (cf.\ also \cite{Pyr}). Also the volume of four-dimensional
slices parallel to the uncompactified dimensions of the stacks of branes will
decrease towards the hidden sector. It has been observed that in order to
obtain a viable phenomenology one has to assume that the $E_8$ on the
hidden sector is also broken to a smaller group \cite{Nilles}.

In order to make connection of these M-theory setups to the Randall-Sundrum
two-brane model, one identifies the visible stack of branes with the
$\Sigma_+$-brane and the hidden stack of branes with the $\Sigma_-$-brane: the
orbifolded coordinate of M-theory is identified with orbifolded bulk
coordinate of the RS model, and the Calabi-Yau space is neglected in the
effective five-dimensional description due to its small volume \cite{RS1}.
The major deviation of the RS model from M-theory ideas is that the size of
the orbifolded bulk may be large. In M-theory models the scale factor for the
uncompactified coordinates corresponding to the warp factor $a(y)$ in the RS
model is tied to the Calabi-Yau volume. A decreasing warp factor also leads to
a decrease of the Calabi-Yau volume. In the M-theory solution of \cite{LOSW}
the relation between the four-dimensional scale factor and the Calabi-Yau
volume reads, for example,
\begin{equation}
V(y) \sim a^6 (y) \, ,
\end{equation}
where $y$ denotes the coordinate of the orbifolded dimension. In scenarios
with a large orbifolded dimension, the dependence of the Calabi-Yau volume on
the scale factor should be weaker because the string
scale sets a lower limit for the size of the Calabi-Yau manifold.
Nevertheless, if M-theory can be extrapolated to setups with a large bulk, one
has to expect a considerable decrease of the Calabi-Yau volume from the
visible to the hidden brane. The crucial observation for our argument is now
that the vacuum expectation value of the gaugino condensate $\eta$ located on
the hidden brane depends strongly on the inverse size of the Calabi-Yau space
\cite{gaugino},
\begin{equation}
\eta \sim \frac{\alpha^3}{V^{9/2}} 
\exp\left[ - 18\pi \frac{V}{b_0 \alpha}(S-\beta T) \right] \, .
\label{condens}
\end{equation}
Here, $\alpha$ is the 10-dimensional gauge-coupling constant, $b_0$ is the
beta-coefficient of the gauge-coupling renormalization group, $S$ and $T$ are
moduli of the Calabi-Yau space and $\beta$ describes loop-corrections to the
moduli fields.  From the structure of (\ref{condens}) we find that even a
moderate decrease of $V$ with the warp factor leads to a strong increase of
the vacuum expectation value of the gaugino condensate, which will therefore be
very large in a RS-like model deduced from M-theory.

Turning now to the theory of cosmic strings, it is well known that the
quadrupole moment of a cosmic string is proportional to the square of the 
vacuum expectation value $\eta$ of the corresponding symmetry breaking,
\begin{equation}
\mu \sim \Gamma \eta^2 / \omega^{3} \, ,
\end{equation}
where $\Gamma \approx 50 \ldots 100$ is a numerical coefficient depending on
the trajectory and shape of the string loop and $\omega$ is the characteristic
frequency of string oscillations \cite{VilShel}. Therefore a large gaugino
vacuum expectation value $\eta$ would lead to a very large quadrupole moment
for cosmic strings that are produced during the breaking of the hidden $E_8$, which
in turn will generate high-amplitude gravitational waves on the hidden brane.
The described effect may easily compensate the damping factor (\ref{damping})
and lead to strong gravitational waves on the hidden brane
and, as a consequence, also on the
$\Sigma_+$-brane.  These in turn can be distinguished from waves
originating from the $\Sigma_+$-brane by detectable RIGO's.

Note that the described mechanism is not spoiled by the dependence of the
effective gravitational coupling constant on the volume of the Calabi-Yau
space, $G_5 \sim G_{11}/V$, because the dependence of the gaugino vacuum
expectation value on the size of the Calabi-Yau space is much stronger
than that of the effective five-dimensional gravitational coupling. 

At first glance our result about a strongly enhanced symmetry-breaking vacuum
expectation value on the hidden brane compared to the visible brane seems to
be in sharp contrast to the conventional interpretation of the solution to the
hierarchy problem in the RS-model (cf.\ e.\,g.\ \cite{RS1,LR,PomarolGNS}).
This conventional interpretation states that the vacuum expectation value of a
symmetry breaking on the hidden brane should be suppressed by two powers of
the warp factor compared to the same value on the visible brane,
\begin{equation}
\eta_- = a^2 \eta_+ \, .
\end{equation}
However, already Randall and Sundrum pointed out that the generation of the
hierarchy could equally well be described by considering the gravitational
scale on the $\Sigma_-$-brane as the derived scale, whereas the symmetry
breaking scale on both branes is to be taken the same \cite{RS1}. That this
interpretation is in fact the correct one has first been demonstrated in
\cite{Grinstein} by considering the fall-off properties of on-brane Green
functions in the Euclidean domain and has been confirmed by an analysis of the
measuring process for masses on the $\Sigma_-$-brane in \cite{KubVol}.
Therefore, all effects of the hierarchy generation in the RS model are already
contained in the damping factor (\ref{damping}) and there is no geometrical
suppression of symmetry-breaking scales on the hidden brane in the RS model.

\subsection{Graviton oscillations on the $\Sigma_-$-brane}\label{RIGO-}

Graviton oscillations on the $\Sigma_-$-brane can be treated along the same
lines as RIGO's on the $\Sigma_+$-brane have been treated in Secs.\ \ref{GW+}
and \ref{RIGO+}. Abandoning once again the matrix notation we find for $h^-$
with the help of the spectral representation (\ref{MLG}) the equations of
motion
      \begin{equation} 
       h^-= -\left.\frac{16\pi G_4}{\Box} 
       \right|_{\rm ret} 
(a^2 T^+ +a^4 T^-) 
        - \sum_{i=1} \left.\frac{16\pi G_4}{\Box-m_i^2} 
        \right|_{\rm ret} 
\left( - \frac {a^2}{\mathcal{J}_2}T^+ + a^2 T^-\right). 
\end{equation}
Considering the same set of sources (\ref{source}) we now study the
gravitational waves seen by an observer on the negative tension brane,
\begin{align}
h^-[\,T^+] & = A a^2\, e^{-i\omega t}
  \left( e^{i\omega r} 
   - \sum_{i=1} \frac{1}{J_2[lm_i/a]} e^{i \sqrt{\omega^2-m_i^2}r} \right) 
    \, ,
\\
h^-[\,T^-] & = A a^2\, e^{-i\omega t}
 \left(a^2 e^{i\omega r} + \sum_{i=1} e^{i \sqrt{\omega^2-m_i^2}r} \right)  
    \, .
\end{align}
For simplicity we suppose again that only the massless and the first massive
modes are generated, i.\,e. $m_1 < \omega < m_2$. Then, for the waves
generated by the source on the $\Sigma_+$-brane and the $\Sigma_-$-brane 
we have, respectively,
 \begin{flalign} 
 h^-[\,T^+] & = Aa^2\, e^{-i\omega t} 
 \left(e^{i\omega r} -  \frac 1{\mathcal{J}_2}  
 e^{i \sqrt{\omega^2-m_1^2}r} \right) \, ,       \label{h-of+} 
\\  
 h^-[\,T^-] & = A a^2\, e^{-i\omega t}  
 \left(a^2 e^{i\omega r} + e^{i \sqrt{\omega^2-m_1^2}r} \right) \, ,
\label{h-of-} 
 \end{flalign} 
where  
$A=4\,G_4\,\mu/r$ is the amplitude of the massless mode generated on  
the $\Sigma_+$-brane.  The 
amplitudes detected by a gravitational-wave interferometer are given by the 
absolute values of (\ref{h-of+}) and (\ref{h-of-}), 
\begin{align} 
\big|h^-[\,T^+]\big| &= \mathcal{A}^+ 
\left[ 1+ \frac{4 \mathcal{J}_2}{(\mathcal{J}_2 -1)^2}  
\sin^2 \left( \frac{\pi r}{L}\right) 
\right]^{1/2}, 
\label{absh-of+} 
\\ 
\big|h^-[\,T^-]\big| &=  
\mathcal{A}^- 
\left[ 1 - \frac{4 a^2}{(1+a^2)^2}  
 \sin^2 \left( \frac{\pi r}{L}\right) 
\right]^{1/2}. 
\label{absh-of-} 
\end{align} 
Here $L$, the oscillation length of the amplitude modulation of the 
gravitational wave, is given by (\ref{length}).  The pre-factors
of the amplitudes (\ref{absh-of+}) and (\ref{absh-of-}) are
\begin{align} 
\mathcal{A}^+ &= \left(\frac 1{\mathcal{J}_2} -1\right) a^2 A
\approx 1.5\, a^2 A \, ,
\\ 
\mathcal{A}^- &= ( 1 + a^2 ) a^2 A
\approx a^2 A \, . 
\end{align} 
Therefore, for the ratio of the typical intensities measured on the
$\Sigma_-$-brane we have,
\begin{equation}
\frac{\mathcal{A}^+}{\mathcal{A}^-} \approx 1.5 \, .
\label{I-ratio}
\end{equation}
For waves on the $\Sigma_-$-brane we thus find that the intensities of waves
produced on this brane and on the other brane are of the same 
order of magnitude.
Therefore at first glance the situation of an observer on the $\Sigma_-$-brane
seems to be favorable for observing graviton oscillations.

Unfortunately, this effect is spoiled by the dependence of the graviton mass
--- and therefore of the oscillation length --- on the position of the brane.
In order to find the oscillation length actually measured by an observer on
the brane, we have to use the coordinate system in which an observer does
his measurements. The coordinates of an observer will always be chosen with
respect to the Lorentz metric $\eta_{\mu\nu}$ (cf.\ \cite{KubVol}). 
No rescaling is necessary for studying effects on the $\Sigma_+$-brane
because there the induced metric coincides with the Minkowski metric.
On the $\Sigma_-$-brane
the ``physical'' coordinates used by an observer are related to those given
with respect to the induced metric $g_{\mu\nu} = a^2 \eta_{\mu\nu}$ by
\begin{equation}
x^\mu_\text{phys} = a x^\mu.
\end{equation}
In the equations of motion (\ref{eom}), the flat space d'Alembertian has to be
replaced by the one with respect to the physical coordinates,
\begin{equation}
\Box = \eta^{\mu\nu} 
\frac{\partial }{\partial x^\mu}\frac{\partial }{\partial x^\nu}
 = a^2 \,
\eta^{\mu\nu}
\frac{\partial }{\partial x_\text{phys} ^\mu }
\frac{\partial }{\partial x_\text{phys}^\nu}
 = a^2 \Box_\text{phys}.
\end{equation}
To be precise, we would also have to rescale the metric perturbation
$h_{\mu\nu}$ (with respect to the background metric $g_{\mu\nu}$) to the
measured one $h^\text{phys}_{\mu\nu} = h_{\mu\nu}/a^2(y)$ (with respect to the
Lorentz metric). We will not perform this step because we are comparing
amplitudes measured on the same brane. Therefore this rescaling does not
change the ratio between the amplitudes. In physical coordinates, the
source-free equations of motion on the $\Sigma_-$-brane are given by
\begin{align}
- \Box_\text{phys} {h^0}^\text{phys}_{\mu\nu} =0,
\quad \quad
- \left[ \Box_\text{phys} - \frac{m_i^2}{a^2} \right] 
{h^i}^\text{phys}_{\mu\nu} =0,
\end{align}
whereas the equations of motion on the $\Sigma_+$-brane remain unchanged.
Therefore the graviton mass appears different when observed on a different
brane,
\begin{equation}
m_i^+ = m_i = j_i \frac al,
\quad\quad
m_i^- = \frac {m_i}a = \frac {j_i}l.
\end{equation}
This rescaling is characteristic of bulk fields in the Randall-Sundrum
two-brane model \cite{KubVol,Grinstein}.\footnote{Note that the rescaling
  does not apply to on-brane fields and therefore does not affect the symmetry
  breaking scale considered in Sec.\ \ref{Mtheory} (cf.\ 
  \cite{KubVol,Grinstein}).}

Whereas the measured graviton masses $m_i^+$ on the $\Sigma_+$-brane depend
on the warp factor, the masses $m_i^-$ on the $\Sigma_-$-brane 
are independent of
the inter-brane distance. From these masses we
can derive the oscillation length observed on the branes for a GW with a 
wave-number $\omega \gtrsim m_i^-$. In this limit we have
\begin{equation}
L = \frac{2\pi}{m_1^-}.
\end{equation}
On the $\Sigma_-$-brane this becomes
\begin{equation}
L^- = \frac{2\pi}{m_1^-} = 2\pi \frac l{j_1}.
\end{equation}
By means of this expression, the upper experimental limit 
on the AdS radius $l \lesssim
10^{-4} \,{\rm m}$ also yields the upper limit on the oscillation length on
the $\Sigma_-$-brane. However, an oscillation length of sub-millimeter size is
clearly unobservable.  Therefore graviton oscillations will be hidden to an
observer on the $\Sigma_-$-brane. 

To conclude, the situation is unfavorable for the detection of RIGO's by an
observer on the $\Sigma_-$-brane: albeit being connected with strong GW's, the
observer cannot detect the oscillations because their oscillation length is
far too short to exhibit itself in interferometric detectors.

\section{RIGO's in bi-gravity models}\label{bigrav}

Bi-gravity models are braneworld setups in which the effective
four-dimensional gravity on the brane which is considered as our Universe is
mediated by one massless and one light massive gravitational mode which have
approximately equally strong couplings to matter on the visible brane. The rest
of the KK spectrum is separated by a mass gap and its coupling to matter on
the visible brane is strongly suppressed. Meanwhile there are various
realizations of these setups \cite{bigrav,KMP}, although only the
six-dimensional ones do not seem to suffer from a VvDZ discontinuity for the
massive mode or from a phenomenologically unrealistic negative curvature of
the visible brane.  For our consideration we do not need to consider a specific
realization of these models but can content ourselves with the general
structure of their equations of motion and the assumption that phenomenological
constraints from the VvDZ discontinuity are somehow circumvented. Then the
four-dimensional effective equations of motion for the transverse-traceless
sector of a generic bi-gravity model reads
\begin{equation}
h^\text{vis} =
- 16 \pi \widetilde{G} \left[
\frac{c_0}{\Box_\text{ret}} + \frac{c_1}{\Box_\text{ret} - m_1^2} 
+ \sum_{n=2} \frac{c_n}{\Box_\text{ret} - m_n^2}
\right] T^\text{vis} \, ,
\end{equation}
where
\begin{align}
&c_0 \approx c_1 \, , \quad \quad 
&&c_n \ll c_0 \, , \quad&& n \geq 2,& & \label{couplings} \\
&m_1 \ll \Delta m_n \, , \quad 
&&\Delta m_n = m_n - m_{n-1} \, ; \quad&& n \geq 2 & &
\end{align}
we have only included sources $T^\text{vis}$ on the visible brane. 
The effective
gravitational constant $\widetilde{G}$ does not coincide with the usual
effective four-dimensional gravitational constant. Rather we have
\begin{equation}
G_4 \approx \widetilde{G} (c_0 + c_1) \, .
\end{equation}

From (\ref{couplings}) we infer that we have a strong mixing between the
massless mode and the first KK mode. This would make bi-gravity models a
natural candidate for strong RIGO's. However, this expectation is spoiled by
phenomenological constraints on this class of models. The major experimental
constraint is established by precision measurements of Newton's law on the
orbital motion of the planets. Our treatment will closely follow that of
\cite{Willmassive} where a purely massive gravitational interaction is
considered.  It is sufficient to consider only the influence of the first KK
mode, since the heavier modes will only contribute to the short-range dynamics.
The non-relativistic static gravitational potential $V(r) = - G_4 M/r$ then 
gets modified by a Yukawa potential to the form 
\begin{equation}
V(r) = - \frac{\widetilde{G}M}r 
\left[c_0 + c_1 \exp \left(- \frac r {\lambda_1}\right) \right]
\, ,
\end{equation}
where $M$ is the mass of the central body.
The gravitational acceleration of a test body in the modified potential 
is ${\bf g} = - {\bf e}_r \, \mu (r)/r^2$
where ${\bf e}_r$ is the unit-vector pointing from the central body to the
test body and
\begin{align}
\mu (r) &= \widetilde{G} M \left[ c_0 + c_1
\left( 1 + r /\lambda_1 \right) \exp \left(- r/\lambda_1 \right)
\right] \nonumber \\
&= \widetilde{G} M \left[ c_0 + c_1
-\frac{c_1}2\left(\frac r{\lambda_1}\right)^2
+ O \left(\frac r{\lambda_1}\right)^3
\right] \, . \label{Yukawa-mu}
\end{align}
For pure Newtonian gravity $\mu =G_4M=$~constant. Its value can be
determined to a high precision from the orbit of the earth around the sun. If
a massive gravitational mode contributes, the $\mu$ determined for the orbits
of other planets will differ from the value
of the earth orbit. For a planet with a
semi-major axis $a_p$ and a period $T_p$, Kepler's third law yields
$\mu^{1/3}(a_p) = a_p^{2/3}(2\pi/T_p)$. Therefore it is convenient to
parametrize the deviation of a planet's $\mu(a_p)$ from the one of the earth
$\mu(a_{\oplus})$ through a small parameter $\eta_p$,
\begin{equation}
1 + \eta_p \equiv
\left( \frac{\mu(a_p)}{\mu(a_{\oplus})} \right)^{1/3} \, .
\label{devpara}
\end{equation}
Using (\ref{Yukawa-mu}) in (\ref{devpara}), we obtain a lower bound on the
wavelength of the first KK mode in terms of the experimental limits on
$\eta_p$,
\begin{equation}
\lambda_1
= \left( \frac{c_1 [a_{\oplus}^2 - a_p^2]}{6 \eta_p (c_0 + c_1)} \right)^{1/2} \, .
\end{equation}
Restricting to the case with $c_0 \approx c_1 \approx 1/2$, we find
\begin{equation}
\lambda_1
= \left( \frac{a_{\oplus}^2 - a_p^2}{12 \eta_p} \right)^{1/2} \, .
\end{equation}
The most restrictive bound on $\eta_p$ comes from the measurement of the Mars
orbit for which we have $\eta_m < - 6.5 \times 10^{-10}$ \cite{Talmadge}. 
This leads to the requirement
\begin{equation}
m_1 \equiv 2 \pi/\lambda_1 < \pi/(10^{12} \, {\rm km})
\label{bigram} 
\end{equation}
for a realistic bi-gravity model. One might try to infer more stringent upper
bounds on the mass of the first KK mode by considering the motions of galaxies
and clusters of galaxies. 
However, these considerations rely heavily on the assumed
amount of dark matter in the universe. In view of the fact that the cosmology
of braneworld models is still in its infancy and can provide us with quite
surprising models of dark matter (see e.\,g. \cite{Mathews}), we resist from
using such constraints which would lead us to an upper mass for the first KK
mode close to the bounds on the graviton mass given by the Particle Data
Group, $m_g \lesssim (10^{19}\, {\rm km})^{-1}$ \cite{PDG}.

However, even the less restrictive solar-system limit (\ref{bigram}) on $m_1$
renders RIGO's in bi-gravity models unobservable. Inserting (\ref{bigram})
into the formula for the oscillation length (\ref{long-length}), we find even
for GW's of a frequency of $10^{-4}\,$Hz, the lowest frequency observable by
LISA, an oscillation length of more than 300 light years. An amplitude
modulation of that length will clearly remain undetectable. The limit on the
wavelength of the first KK mode does however still allow for the possibility
of detecting ultralight KK modes by studying the propagation speed of
gravitational waves with LISA (cf.\ \cite{Willmassive}).

In addition to the bounds on the length of RIGO's from solar-system dynamics,
it is interesting to note that some models even predict a minimal oscillation
length beyond the present-day Hubble radius of our Universe. This is in
particular the case in the Kogan-Mouslopoulos-Papazoglou version \cite{KMP} of
the Karch-Randall model \cite{KR} with two AdS$_4$ branes embedded in AdS$_5$.
This model is particularly attractive because the longitudinal polarization of
the KK gravitons is supressed in the propagator by the geometry of the setting.
Therefore, the VvDZ discontinuity is absent \cite{KMP,KRVvDZ}.

In this model, the strongest constraint on the length of RIGO's comes from the
requirement that the observable cosmological constant of our Universe, i.\,e.\ 
the effective cosmological constant $\Lambda_4$ of one of the AdS$_4$ branes,
should be compatible with the observed magnitude of the cosmological constant,
\begin{equation}
|\Lambda_4| \lesssim 10^{-120} \, M_\text{Pl}^4 \, ,
\end{equation}
where $M_\text{Pl}$ denotes the four-dimensional Planck mass. This 
requirement on the geometry of the model is reflected in the coefficients
 of the harmonics of the bulk spacetime and thereby
restricts the mass of the first KK-mode $m_1$ to a fraction of the inverse
size of the Hubble radius of our Universe, 
\begin{equation}
m_1 \sim e^{-135}/(H^{-1}) \, .
\end{equation}
Using the expression for
the lower limit of the oscillation lenght (\ref{short-length}), 
the value for the first KK mass leads to
an oscillation length for RIGO's,
\begin{equation}
L_\text{min} \sim e^{135} H^{-1} \gg H^{-1} \, ,
\end{equation}
which is much larger than the size of the horizon of our Universe and thus
renders RIGO's unobservable {\em in principle} in this particular model.

\section{Conclusions\label{conc}}

We have studied the phenomenon of radion induced graviton oscillations in the
two-brane world.  First, we have established a method to extract the particle
content from the nonlocal braneworld effective action of \cite{we} for the
Randall-Sundrum two-brane model. This method is easily extendable to
arbitrary two-brane models and provides the missing link between nonlocal
braneworld actions, which are specifically suited for the treatment of
cosmological problems \cite{we,Muko,Andrei}, and spectral representations of the
effective action particularly suited to particle-phenomenology considerations
 \cite{KubVol1,KubVol}.

From the equations of motion for the graviton sector, we have found a mixing of
massless and massive modes which depends parametrically on the radion.  This
mixing leads to the effect of radion induced graviton oscillations which is
in principle observable.  RIGO's are a feature of every higher-dimensional
spacetime model, since there will always occur amplitude modulations in GW's
which are a mixture of a massless mode and KK modes. However, in traditional
models with flat extra dimensions, the mass of the first KK mode is so big
that it will neither be produced by astrophysical sources nor lead to
oscillation lengths of macroscopic size. In contrast to this, warped
geometries allow KK mode masses which are so low that they can lead to
oscillations of detectable length. In particular, waves from sources on the
hidden brane show strong oscillations on the visible brane. The amplitudes of
these waves are strongly suppressed by the warped geometry and thus, at
first sight, seem to remain undetectable. However, by using simple M-theory
motivated scaling arguments, we have demonstrated that one should expect a
network of cosmic strings on the hidden brane which would produce a background
of high amplitude gravitational waves. The M-theory scaling properties
considered may easily compensate the geometrical damping of GW's from the
hidden brane.  Therefore GW's with RIGO's stemming from the hidden brane may
actually become a relevant effect for gravitational-wave astronomy. The
characteristic pattern of RIGO's may even help to discriminate between
noise and the stochastic GW background in GW interferometers.

At first sight another natural candidate for observable RIGO's are bi-gravity
models. In these models the massless graviton and the lightest massive one are
coupled to matter with nearly equal strength and, thus, produce strong
oscillations.  Unfortunately, bi-gravity models which exhibit
RIGO's of experimentally detectable oscillation-length are already ruled out
by precision measurements of Newton's law on solar-system scales.

The effects connected with the mixing and
oscillations of quantum states in a multidimensional spacetime
have received some attention in the literature
\cite{brane-osc,brane-osc1,brane-osc2,CP}.
Neutrino mixing and oscillations were reconsidered in
\cite{brane-osc,brane-osc1,brane-osc2}, while the mixing between
quarks and an attempt
to explain the origin of CP violation was described in
\cite{CP}. Nevertheless, to our knowledge, the possibility of oscillations
between different graviton states and their potential observability in GW
interferometers has not been considered before.

\section*{Acknowledgements}

A.O.B. and A.Yu.K. are grateful for the hospitality of the Theoretical Physics
Institute, University of Cologne, where a major part of this work has been
done due to the support of the DFG grant 436 RUS 113/333/0-2. A.O.B. is also
grateful to the University of Munich where the final stage of this work was
done under the grant SFB 375.  A.Yu.K.\ is grateful to CARIPLO Science
Foundation.  The work of A.O.B.\ was also supported by the RFBR grant
02-01-00930 and the grant of Leading Scientific Schools 1578.2003.2, while
A.Yu.K.\ was supported by the RFBR grant No 02-02-16817 and 
by the scientific school grant No 2338.2003.2 of the Russian Ministry
of Science and Technology.
A.R.\ was supported by the DFG-Graduiertenkolleg ``Nonlinear Differential
Equations''.

\begin{appendix}
\section{Diagonalization of the kinetic and mass terms}

Here we describe in detail the diagonalization of the kinetic and massive
operators and the calculation of the mixing parameters from Sec.\ \ref{diag}.
The procedure of the diagonalization of the matrices ${\bf D}$ and ${\bf M}$
will include three successive operations.  First we shall find the eigenvalues
$\lambda_1,\lambda_2$ of the operator ${\bf D}$ and shall construct an
orthogonal matrix $O$ which provides the transition from an old basis to the
basis of normalized eigenvectors of the operator ${\bf D}$. Application of the
transformation $O$ to the matrix ${\bf D}$ transforms it to a diagonal matrix
whose elements coincide with the eigenvalues of ${\bf D}$,
\begin{equation}
O{\bf D}O^T = \left[\begin{array}{cc}\lambda_1&0\\
0&\lambda_2\end{array}\right]. \label{stage1}
\end{equation}

Secondly, in order to transform this matrix into the unit matrix, we
act on the eigenvectors by the generalized dilatation matrix $\Delta$,
\begin{equation}
\Delta = \left[\begin{array}{cc}\sqrt{\lambda_1}&0\\
0&\sqrt{\lambda_2}\end{array}\right]. \label{dilat}
\end{equation}
Of course, the action of the inverse operators $\Delta^{-1}$ on the matrix (\ref{stage1})
yields the unit matrix,
\begin{equation}
\Delta^{-1}O\,{\bf D}\,O^T\Delta^{-1} = \left[\begin{array}{ccc}1&&0\\
0&&1\end{array}\right]. \label{dilat1}
\end{equation}

The simultaneous action of the operators $O$ and $\Delta$ on the matrix ${\bf
  M}$ (\ref{oper-gr1}) transforms it into
\begin{equation}
\tilde{{\bf M}} = \Delta^{-1}O\,{\bf M}\,O^T\Delta^{-1}, \label{mass}
\end{equation}
which is still degenerate. The non-zero eigenvalue of this matrix, which
gives the squared mass of the massive graviton, is equal to the
trace of
this matrix,
\begin{equation}
M^2 = {\rm Tr}\tilde{{\bf M}}. \label{mass1}
\end{equation}

As a third step, in order to finally obtain an expression for the
transformation from the initial gravitational modes $h^+$ and $h^-$ to the new
modes $h_M$ and $h_0$ corresponding to the massive and massless gravitons, we
should use another orthogonal rotation $Q$, diagonalizing the matrix
$\tilde{{\bf M}}$ in such a way that its diagonal elements coincide with its
eigenvalues, i.\,e.\ 
\begin{equation}
Q\tilde{{\bf M}}Q^T = \left[\begin{array}{cc}M^2&0\\
0&0\end{array}\right]. \label{mass2}
\end{equation}
Of course, the application of the orthogonal transformation $Q$ to the
transformed kinetic matrix (\ref{dilat1}) does not change it because it is
a unit matrix.

Thus, one can formulate the transformation between the modes $h^+$ and
$h^-$ and the eigenmodes of the operator (\ref{oper-gr}) in the
form
\begin{equation}
\left[\begin{array}{c}h_M\\h_0\end{array}\right]
= Q \Delta O \left[\begin{array}{c}h^+\\h^-\end{array}\right].
\label{mixing}
\end{equation}
The inverse transformation is
\begin{equation}
\left[\begin{array}{c}h^+\\h^-\end{array}\right]
= O^T \Delta^{-1} Q^T \left[\begin{array}{c}h_M\\h_0\end{array}\right].
\label{mixing1}
\end{equation}
The transformations (\ref{mixing}) and
(\ref{mixing1}) are not orthogonal. This is connected with the fact
that we had to apply the generalized dilatation matrix $\Delta$ to normalize
properly the kinetic part of the effective action.

The eigenvalue equation for the kinetic matrix ${\bf D}$, defined in
Eq.~(\ref{oper-gr2}), has the following form,
\begin{equation}
\lambda^2 - \lambda\alpha(a^4 + 6a^2 + 1) +
3\alpha^2 a^2(a^2+1)^2 = 0 \, , \label{eigenvalue}
\end{equation}
where we have introduced the abbreviation 
$\alpha = (1-a^2)/3a^2(1+a^2)^2$.
The solutions are
\begin{equation}
\lambda_{1,2} = \frac{\alpha(a^4+6a^2+1)}{2} \pm
\frac{\alpha\sqrt{a^8+14a^4+1}}{2}. \label{eigenvalue1}
\end{equation}
Now, we substitute these eigenvalues into an eigenvector equation,
\begin{equation}
({\bf D} - \lambda_{1,2} {\bf I})\psi_{1,2} = 0 \, ,
\label{eigenvector}
\end{equation}
where ${\bf I}$ is a unit matrix and $\psi_{1,2}$ are eigenvectors
corresponding to the eigenvalues $\lambda_{1,2}$, respectively.
We choose the normalized eigenvectors
\begin{equation}
\psi_1 = \frac{1}{N_1}\left[\begin{array}{c}
D_{12}\\
\lambda_1-D_{11}\end{array}\right],
\label{eigenvector1}
\end{equation}
\begin{equation}
\psi_2 = \frac{1}{N_2}\left[\begin{array}{c}
D_{12}\\
\lambda_2-D_{11}\end{array}\right],
\label{eigenvector2}
\end{equation}
where $D_{ij}$ are the corresponding elements of the matrix ${\bf D}$,
while the normalization factors are equal to
\begin{align}
N_1 &= \sqrt{D_{12}^2 + (D_{11}-\lambda_1)^2},\nonumber\\
N_2 &= \sqrt{D_{12}^2 + (D_{11}-\lambda_2)^2}. \label{norm}
\end{align}
Correspondingly, the orthogonal matrix providing the transformation from the
old basis $\left[\begin{array}{c}1\\0\end{array}\right],
\left[\begin{array}{c}0\\1\end{array}\right]$ to the new basis $\psi_1,
\psi_2$ (given by the expressions (\ref{eigenvector1}), (\ref{eigenvector2})),
has the form
\begin{equation}
O = \left[\begin{array}{cc}\displaystyle\frac{D_{12}}{N_1}&
\displaystyle\frac{D_{12}}{N_2}\\
\displaystyle\frac{\lambda_1-D_{11}}{N_1}&
\displaystyle\frac{\lambda_2-D_{11}}{N_2}\end{array}\right].
\label{orthog}
\end{equation}

Using the expressions for the orthogonal matrix $O$ and for the generalized
dilatation matrix $\Delta$, one can get an expression for the rotated matrix
$\tilde{\bf M}$ by substituting Eqs.~(\ref{orthog}) and (\ref{dilat}) into
Eq.~(\ref{mass}),
\begin{equation}
\tilde{{\bf M}} = \Delta^{-1}O{\bf M}O^T\Delta^{-1}
= \beta\left[\begin{array}{cc}
\displaystyle\frac{\tilde{M}^2}{\lambda_1}&
\displaystyle\frac{\tilde{M}\tilde{m}}{\sqrt{\lambda_1\lambda_2}}\\
\displaystyle\frac{\tilde{M}\tilde{m}}{\sqrt{\lambda_1\lambda_2}}&
\displaystyle\frac{\tilde{m}^2}{\lambda_2}\end{array}\right], \label{mass3}
\end{equation}
where
\begin{equation}
\tilde{M} = a^2 D_{12}\left(\frac{a^2}{N_1} - \frac{1}{N_2}\right),
\label{mass4}
\end{equation}
\begin{equation}
\tilde{m} = \frac{a^2(r-s)}{N_1} + \frac{r+s}{N_2}, \label{mass5}
\end{equation}
and
\begin{equation}
r = \frac{\lambda_1-\lambda_2}{2}\, ,\quad  \quad
s = \frac{D_{11} - D_{22}}{2} \, . \label{mass6}
\end{equation}

Now we are in a position to calculate the value of the mass of the massive
graviton mode by substituting the formulas (\ref{mass3}) -- (\ref{mass6}),
together with formula (\ref{eigenvalue1}) and the explicit values of the matrix
elements $D_{ij}$ from formula (\ref{oper-gr}), into Eq.~(\ref{mass1}).
Straightforward but rather cumbersome calculations lead us to the simple
expression
\begin{equation}
M^2 = \frac{24 a^2 (1+a^2)}{l^2 (1-a^2)^2}. \label{Mass}
\end{equation}

To accomplish the task of simultaneous diagonalization of the matrices
${\bf D}$ and ${\bf M}$, we should find the matrix $Q$ rotating the matrix
$\tilde{\bf M}$ to the diagonal form (\ref{mass2}).  One can find normalized
eigenvectors of the matrix $\tilde{\bf M}$ by solving the equations
\begin{align}
(\tilde{\bf M} - M^2 {\bf I})\phi_1 &=0 \, ,\nonumber \\
\tilde{\bf M}\phi_2 &=0. \label{eigenvector3}
\end{align}
These eigenvectors have the form
\begin{align}
\phi_1 &= \frac1N\left[\begin{array}{c}\sqrt{\lambda_2}\tilde{M}\\
\sqrt{\lambda_1}\tilde{m}\end{array}\right],\nonumber \\
\phi_2 &= \frac1N\left[\begin{array}{c}\sqrt{\lambda_1}\tilde{m}\\
-\sqrt{\lambda_2}\tilde{M}\end{array}\right], \label{eigenvector4}
\end{align}
where
\begin{equation}
N = \sqrt{\lambda_2 \tilde{M}^2 + \lambda_1\tilde{m}^2}.
\label{eigenvector5}
\end{equation}
Correspondingly, the orthogonal matrix $Q$ reads
\begin{equation}
Q = \frac1N \left[\begin{array}{cc}\sqrt{\lambda_2}\tilde{M}&
\sqrt{\lambda_1}\tilde{m}\\
\sqrt{\lambda_1}\tilde{m}&-\sqrt{\lambda_2}\tilde{M}
\end{array}\right]. \label{Q}
\end{equation}

Substituting the matrices $O,\Delta$ and $Q$, given by Eqs.~(\ref{orthog}),
(\ref{dilat}) and (\ref{Q}), into Eq.~(\ref{mixing}), we find the expressions
(\ref{mixing30}), (\ref{mixing3}), describing the transformation from the old
graviton modes $h^+$ and $h^-$ to the modes $h_M$ and $h_0$, which have already
been given in Sec.~\ref{diag}. Analogously, one can also obtain the
expressions (\ref{mixing40}), (\ref{mixing4}) presented in Sec.~\ref{diag}
and describing the inverse transition from the eigenmodes of the diagonalized
Hamiltonian $h_M$ and $h_0$ to the old graviton modes $h^+$ and $h^-$.
\end{appendix}

\end{document}